\documentclass[onecolumn,12pt]{pasj01}
\usepackage{epsfig}
\usepackage{subfigure}
\usepackage{graphicx}
\usepackage{rotating}
\usepackage{natbib}
\usepackage{booktabs}
\usepackage{bm}
\usepackage{threeparttable}

\newcommand{\mum}{\mu {\rm m}}

\SetRunningHead{Okabe et al.}
{Image Stacking Analysis of SDSS Galaxies 
with AKARI Far-Infrared Surveyor Maps}
\begin{document}

\title{Image Stacking Analysis of SDSS Galaxies with AKARI Far-Infrared Surveyor Maps at 65$\mum$, 90$\mum$, and 140$\mum$}

\author{Taizo \textsc{Okabe}\altaffilmark{1}, 
Toshiya \textsc{Kashiwagi}\altaffilmark{1},
Yasushi \textsc{Suto}\altaffilmark{1,2}, \\
Shuji \textsc{Matsuura}\altaffilmark{3,4}, 
Yasuo \textsc{Doi}\altaffilmark{5}, 
Satoshi \textsc{Takita}\altaffilmark{3} 
and Takafumi \textsc{Ootsubo}\altaffilmark{5},}
\altaffiltext{1}{Department of Physics, The University of Tokyo,
Tokyo 113-0033}
\altaffiltext{2}{Research Center for the Early Universe, School of Science, 
The University of Tokyo, Tokyo 113-0033}
\altaffiltext{3}{Institute of Space and Astronautical Science, 
Japan Aerospace Exploration Agency, Kanagawa 252-5210}
\altaffiltext{4}{present address: Department of Physics, School of Science and Technology,
Kwansei Gakuin University 2-1 Gakuen, Sanda, Hyogo 669-1337, Japan}
\altaffiltext{5}{Department of Earth Science and Astronomy, Graduate School of Arts and Science, 
The University of Tokyo, Tokyo 153-8902}
\email{taizo.okabe@utap.phys.s.u-tokyo.ac.jp}


\maketitle

\begin{abstract}
We perform image stacking analysis of Sloan Digital Sky Survey (SDSS) photometric galaxies over the AKARI Far-Infrared Surveyor (FIS) maps at 65$\mu{\rm m}$, 90$\mu{\rm m}$, and 140$\mu{\rm m}$.  
The resulting image profiles are decomposed into the central galaxy component (single term) and the nearby galaxy component (clustering term), as a function of the $r$-band magnitude, $m_r$ of the central galaxy. 
We find that the mean far-infrared (FIR) flux of a galaxy with magnitude $m_r$ is well fitted with $f^s_{90\mu{\rm m}}=13\times10^{0.306(18-m_r)}$[mJy]. 
The FIR amplitude of the clustering term is consistent with that expected from the angular-correlation function of the SDSS galaxies, but galaxy morphology dependence needs to be taken into account for a more quantitative conclusion. 
We also fit the spectral energy distribution of stacked galaxies at 65$\mu{\rm m}$, 90$\mu{\rm m}$, and 140$\mu{\rm m}$, and derive a mean dust temperature of $\sim 30$K.  
This is consistent with the typical dust temperature of galaxies that are FIR luminous and individually detected.
\end{abstract}

\section{Introduction}

Dust plays an important role in formation and chemical evolution of
galaxies.  Dust absorbs and scatters the ultraviolet (UV) light
associated with star formation activities, and approximately re-emits
half of the star light in infrared (IR) (e.g. \citealt{Lutz2014}).  The
infrared dust emission from individual galaxies, however, is very
difficult to detect in IR generally, except for bright sources.  This is
why image stacking analysis is very useful in measuring and
characterizing the IR emission from typical galaxies in a statistical
fashion.

Recently, for instance, \citet{Kashiwagi2013} (hereafter KYS13) performed
the stacking analysis of SDSS \citep{York2000} photometric galaxies over
the Galactic extinction map by Schlegel, Finkbeiner, and Davis
(\citealt{SFD98}, hereafter SFD). The SFD extinction map is constructed
basically from IRAS/ISSA FIR emission map at 100$\mum$ with dust
temperature correction using the COBE data.

KYS13 were originally motivated to explain the anomaly of the SFD
extinction map \citep{Yahata2007,KashiwagiSFD} in terms of the FIR
emission contamination from galaxies.  Indeed they are successful in
detecting the additional extinction statistically over the SFD map
that should be ascribed to SDSS galaxies.  Any further quantitative
interpretation of the result, however, is limited by the poor angular
resolution (FWHM $\sim 6'$) of the IRAS 100$\mum$ map.

In this paper, we perform the similar stacking analysis using the
Far-Infrared Surveyor (FIS; \citealt{Doi2015}, \citealt{Takita2015})
onboard the Japanese infrared satellite AKARI \citep{Murakami2007}.  FIS
covers almost $98\%$ of the whole sky with spatial resolution of
$\sim{1'.5}$, approximately four times better than that of IRAS.  Also
FIS maps at 65, 90, 140 and 160$\mum$ are useful in estimating the
corresponding dust temperature in a statistical fashion. Indeed the
longer wavelength bands correspond to the peak of dust far-infrared
emission from SDSS photometric galaxies as described below.

The present paper is organized as follows. Section 2 briefly summarizes
the AKARI FIS all-sky map and SDSS DR7 photometric galaxy catalog that
we use throughout the current analysis. Section 3 describes the point
spread functions (PSFs) of FIS, and also presents the method of the
stacking. The stacked images at 90$\mum$ are fitted to the model
profiles in section 4, compared with the prediction on the basis of the
angular-correlation function of SDSS galaxies in section 5.  We repeat
the similar stacking analysis at 65 and 140$\mum$, and derive the dust
temperature of SDSS galaxies in section 6.  Finally \S 7 is devoted to
summary and conclusion of the paper.

\section{Data : SDSS and AKARI FIS}

Our stacking analysis utilizes the two datasets, AKARI FIS
\citep{Kawada2007} and SDSS DR7 data \citep{Abajazian2009}, which we
describe briefly below.

\subsection{AKARI FIS All-Sky Maps}

FIS is an instrument on board the AKARI satellite \citep{Kawada2007}.
It has 4 photometric bands at 65 $\mum$, 90 $\mum$, 140 $\mum$ and 160
$\mum$, and covers almost all the sky.  Figure \ref{fig:sdssregion}
shows the 90$\mum$ image of FIS map in the ecliptic coordinates.  The
absolute calibration of FIS map is based on the comparison with the
COBE/DIRBE data \citep{Takita2015}.  While other component and point
sources are not removed, smooth cloud components of the zodiacal
emission are subtracted from the map following \citet{Gorjian2000}.
Indeed, a faint signature of the residual zodiacal light is visible
around the equator in figure \ref{fig:sdssregion}.

The point-source detection levels of the survey mode at a
signal-to-noise ratio of 5$\sigma$ are 2.4Jy, 0.55Jy, and 1.4Jy at 65
$\mum$, 90 $\mum$ and 140 $\mum$, respectively \citep{Kawada2007}.
\citet{Arimatsu2014} directly measure the point spread functions (PSFs)
of FIS by stacking infrared standard stars of \citet{Cohen1999}.  Their
catalogue includes 422 giant stars with spectral types of K0 to M0, and
\citet{Arimatsu2014} stack 80$\%$ of those stars.  Table
\ref{table:arimatsu} summarizes the values of the PSFs of FIS and the
5$\sigma$ detection limits in each wavelength.  The PSFs of FIS are
elongated along the direction of scanning because of the slow response
of FIS detector.  This is why we show three different values for the
FWHM in Table \ref{table:arimatsu}; along the scan direction (In-scan),
perpendicular to the scan direction (Cross-scan), and their circular
average.  Since the 160$\mum$ data are too noisy, we do not use them
(cf. \citealt{Arimatsu2014}).

\begin{table}[h]
\caption{The FWHMs of the PSFs of AKARI FIS (Table 2 of
\citealt{Arimatsu2014}) and the $5\sigma$ detection limits in each
wavelength.}  
\label{table:arimatsu}
\begin{center}
\begin{tabular}{llll}
\\ \hline \hline
 & 65$\mum$ & 90$\mum$ & 140$\mum$ \\ \hline
In-scan FWHM & $82''$ & $98''$ & $101''$ \\
Cross-scan FWHM & $33''$ & ${55''}$ & ${70''}$ \\
Circular average FWHM & ${53''}$ & ${73''}$ & ${86''}$ \\
5$\sigma$ detection limit & 2.4Jy & 0.55Jy & 1.4Jy \\ \hline
\end{tabular}
\end{center}
\end{table}

\subsection{SDSS DR7 Photometric Galaxies \label{sec:sdss}}

\begin{figure}[t]
\begin{center}
 \FigureFile(100mm,100mm){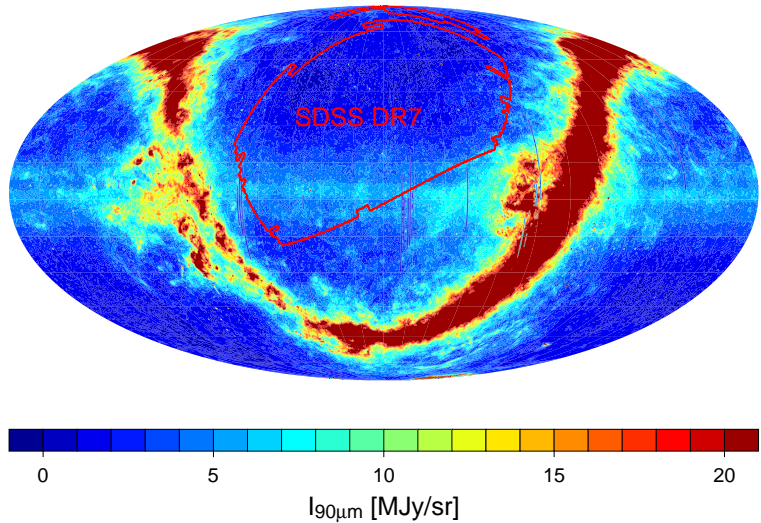}
\end{center}
\caption{All-sky map of 90$\mum$ observed by AKARI in ecliptic
  coordinate. The color scale indicates the intensity of FIR emission
  at 90$\mum$. The region surrounded by the red solid line indicates
  the SDSS DR7 survey region used in the worth Galactic map for the
  present analysis.}
\label{fig:sdssregion}
\end{figure}

Our stacking analysis is based on the SDSS DR7 \citep{Abajazian2009}
photometric galaxy catalog, which covers 11663 ${\rm deg}^2$ of the sky,
with photometry in five passbands: $u, g, r, i,$ and $z$.  (See
\citealt{Stoughton2002, Gunn1998, Gunn2006, Fukugita1996, Hogg2001,
Ivezic2004, Smith2002, Tucker2006, Padmanabhan2008, Pier2003} for more
details of photometric data.)  We use the contiguous regions of
7270$~{\rm deg}^2$ in the north galactic hemisphere (shown in figure
\ref{fig:sdssregion}).  We exclude the masked regions so as to avoid
those objects with unreliable photometry.

Since the spatial distribution of the Galactic stars is likely to be
correlated with that of the Galactic dust, the contamination of stars
in our galaxy sample would affect the interpretation of the stacking
analysis presented below.  Therefore, for ensuring the reliable
star-galaxy separation, we conservatively remove bad photometry data
and fast-moving objects from the ``GALAXY'' sample on the basis of the
SDSS photometry flag; see \citealt{Yahata2007} for more details
of our galaxy sample selection.

Finally, we restrict the magnitude range of our galaxy sample as
$15.5<m_r<20.5$. This is because the star-galaxy separation by the SDSS
photometry pipeline works well for those objects with $m_r<21$
\citep{Yasuda2001}.  Our sample selection mentioned above removes
$24$-$39\%$ of the photometric galaxy candidates in each magnitude bin
(table \ref{table:numbers-of-galaxies}).

\section{Method of Stacking Analysis}

\subsection{Point Spread Function\label{sec:PSF}}

Since the PSFs of AKARI FIS are much larger than the typical size of
SDSS galaxies, each single galaxy on the stacked image is approximated
by the PSFs.  Therefore, it is crucially important to model the PSFs
accurately. In reality, while the PSFs of AKARI FIS are elongated along the
scan directions which align with ecliptic longitudes, we use their
circular averages in the following analysis just for simplicity.  Since
\citet{Arimatsu2014} do not find any systematic dependence of the PSFs
on the fluxes of the sources, we ignore the dependence on the source flux and adopt the same PSF independently of $m_r$ of galaxies.  

\begin{figure*}[t]
\begin{center}
    \FigureFile(55mm,80mm){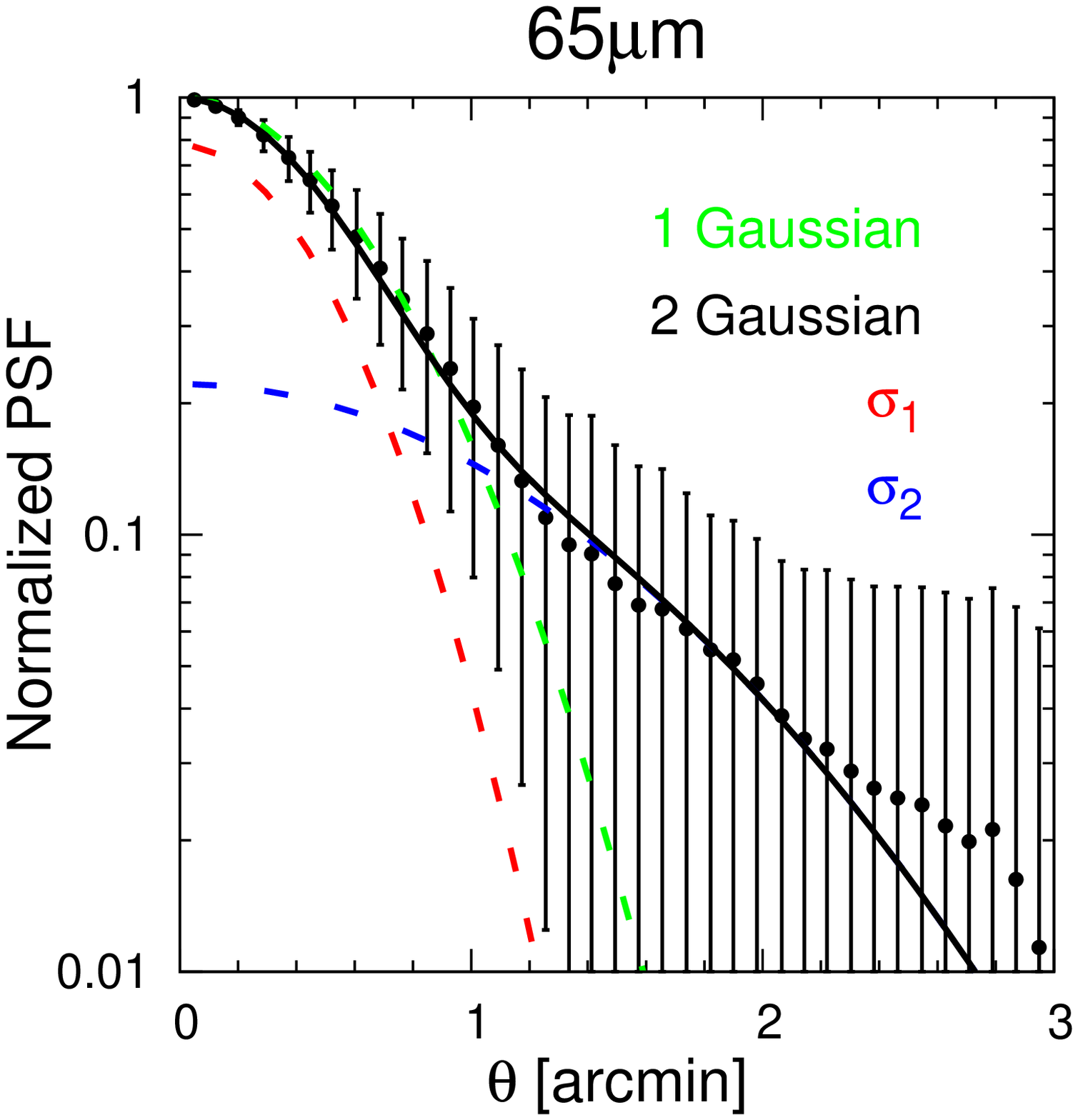}
    \FigureFile(55mm,80mm){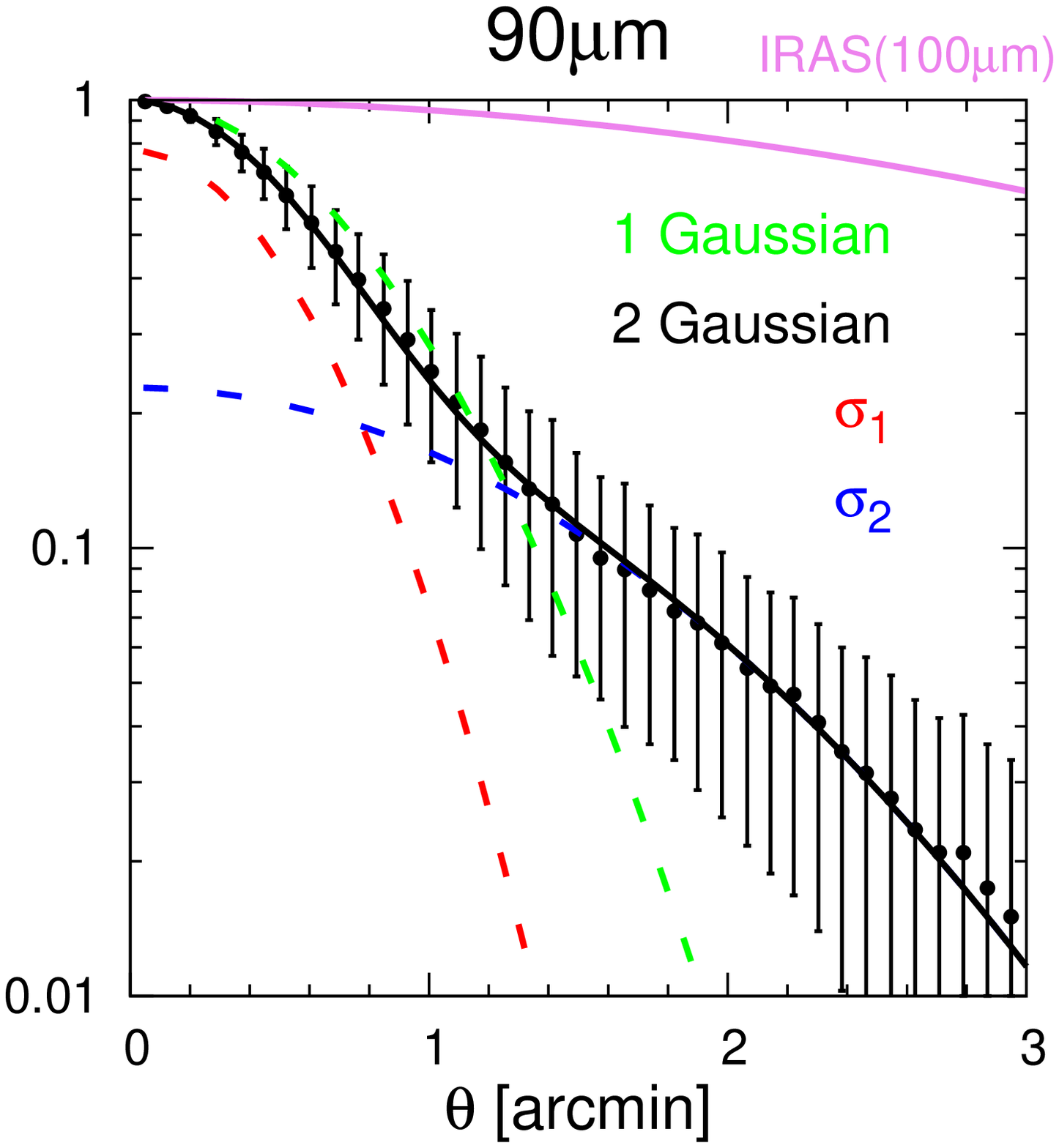}
    \FigureFile(55mm,80mm){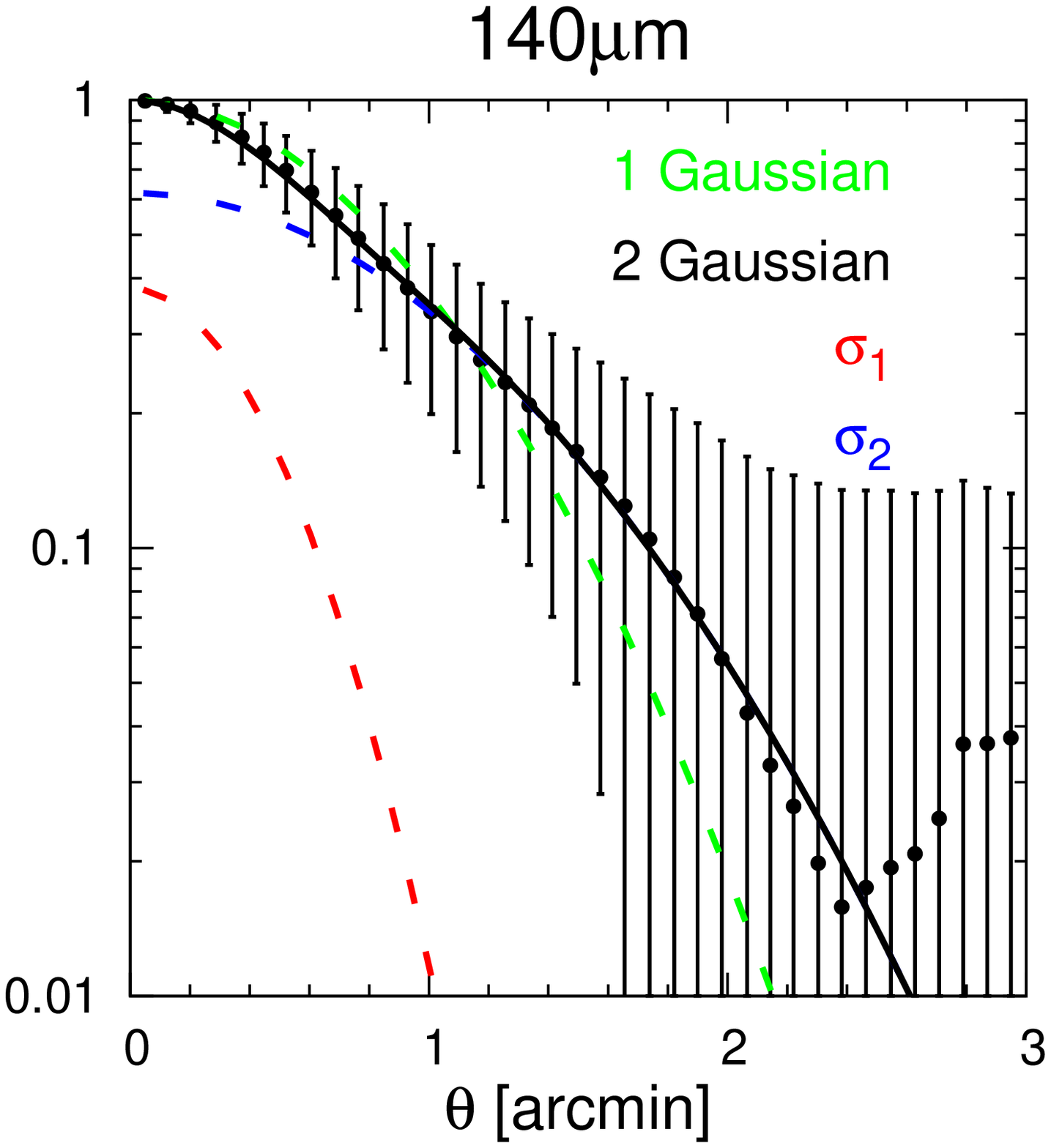}
\end{center}
\caption{Circular-averaged PSFs at 65$\mum$ (left), 90$\mum$ (center),
  and 140$\mum$ (right).  The symbols indicate the PSFs measured by
  \citet{Arimatsu2014}.  The quoted error-bars represents rms in each
  circular bin. The green dashed line and black solid line indicate
  best-fits of single Gaussian and double Gaussian, respectively. The
  red and blue dashed line correspond to two components of double
  Gaussian. The violet solid line in the middle panel represents the
  PSF of IRAS ($100\mum$).}
 \label{fig:PSF}
\end{figure*}

Figure \ref{fig:PSF} shows the radial profiles of the circular-averaged
PSFs.  The quoted error-bars represent rms in each radial bin. The
relatively large rms comes from the anisotropy of the PSFs. Clearly the
circular averaged PSFs have a long tail and are not simply approximated
by a single Gaussian (green dashed lines). Instead, we find that the following
double Gaussian is a good approximation for the PSFs:
\begin{eqnarray}
\label{eq:double-PSF-profile}
W_2(\theta)  &=&
A\exp\left(-\frac{\theta^2}{2\sigma_1^2}\right)
+\left(1-A\right)\exp\left(-\frac{\theta^2}{2\sigma_2^2}\right).
\end{eqnarray}
We fit equation (\ref{eq:double-PSF-profile}) to the circular averaged
PSFs (figure \ref{fig:PSF}) with the three parameters; $\sigma_1,
\sigma_2$ and $A$.  The fit is performed for $\theta<{3'}$ at
65$\mum$ and 90$\mum$, and for $\theta<{2.'4}$ at 140$\mum$
because of the large error bars beyond these scales.  Table
\ref{table:psf} lists the best-fit parameters that correspond to the
black solid lines in figure \ref{fig:PSF}.

\begin{table}[t]
\caption{The number of SDSS galaxies used for the stacking 
analysis, and removed by masking the panels with 
$I_{90\mum}>200$[MJy/sr] for different $r$-band magnitudes.}
\label{table:numbers-of-galaxies}
\begin{center}
\begin{tabular}{lllll}
\hline \hline
$m_r$ & GALAXY & selected & stacked & removed \\
\hline
	$15.5-16.0$ & $56183$ & $34322$ & $34315$ & $7$ \\
	$16.0-16.5$ & $100414$ & $61392$ & $61383$ & $9$  \\
	$16.5-17.0$ & $171767$ & $108553$ & $108514$ & $39$ \\
	$17.0-17.5$ & $291599$ & $194546$ & $194490$ & $56$ \\
	$17.5-18.0$ & $499741$ & $345168$ & $345078$ & 90$$ \\
	$18.0-18.5$ & $861461$ & $617291$ & $617126$ & $165$ \\
	$18.5-19.0$ & $1476678$ & $1099475$ & $1099154$ & $321$ \\
	$19.0-19.5$ & $2511481$ & $1909836$ & $1909289$ & $547$ \\
	$19.5-20.0$ & $4208480$ & $3179658$ & $3178776$ & $882$ \\
	$20.0-20.5$ & $6945931$ & $5042565$ & $5041111$ & $1454$ \\
\hline
\end{tabular}
\end{center}
\end{table}

\begin{table}[h]
\caption{Best fit parameters of the PSF at each passband of FIS.}
\label{table:psf}
\begin{center}
	\begin{tabular}{llll}
	\hline \hline
	 & 65$\mu$m & 90$\mu$m & 140$\mu$m \\
	\hline
	$A$ & $0.78$ & $0.77$ & $0.38$ \\
	$\sigma_{1}$ & ${25''.0}$ & ${27''.7}$ & ${22''.7}$ \\
	$\sigma_{2}$ & ${65''.8}$ & ${73''.8}$ & ${54''.5}$ \\ 
	\hline 
	\end{tabular}
\end{center}
\end{table}

\begin{figure*}[h]
\begin{center}
    \FigureFile(55mm,55mm){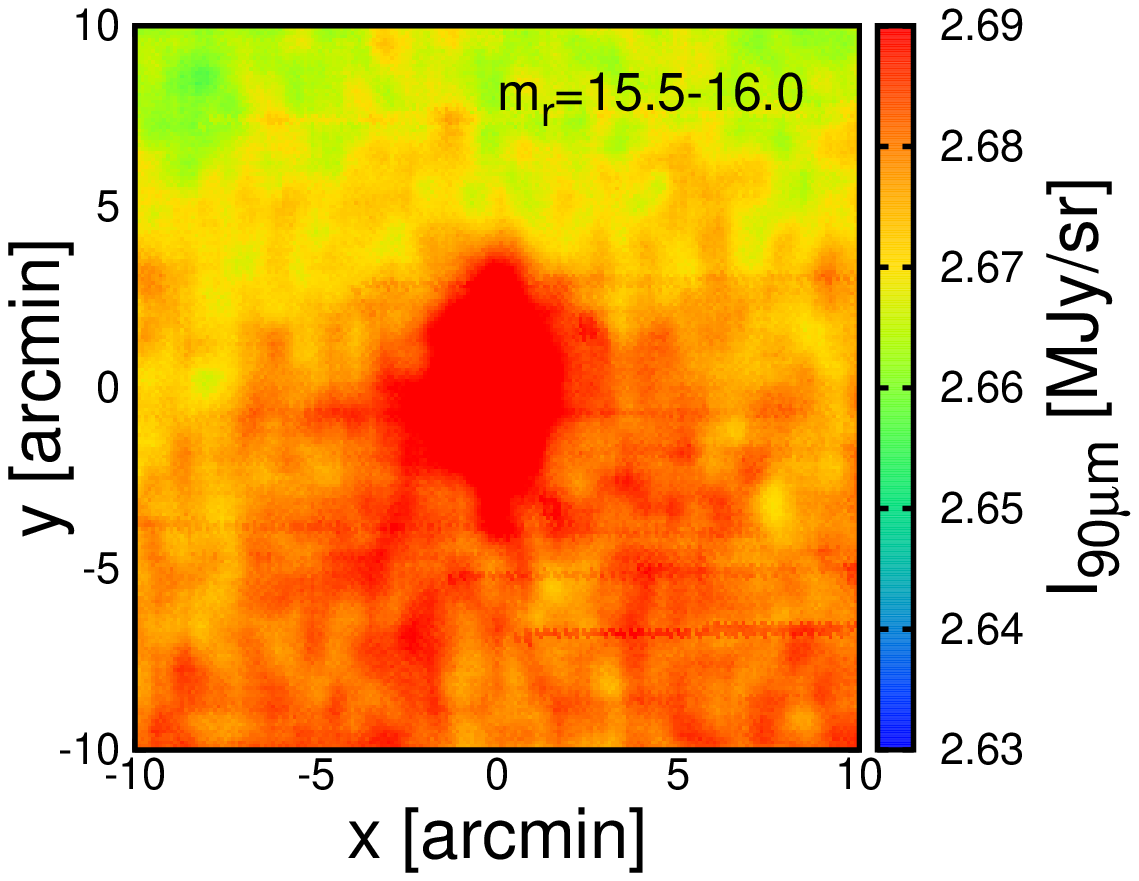}
    \FigureFile(55mm,55mm){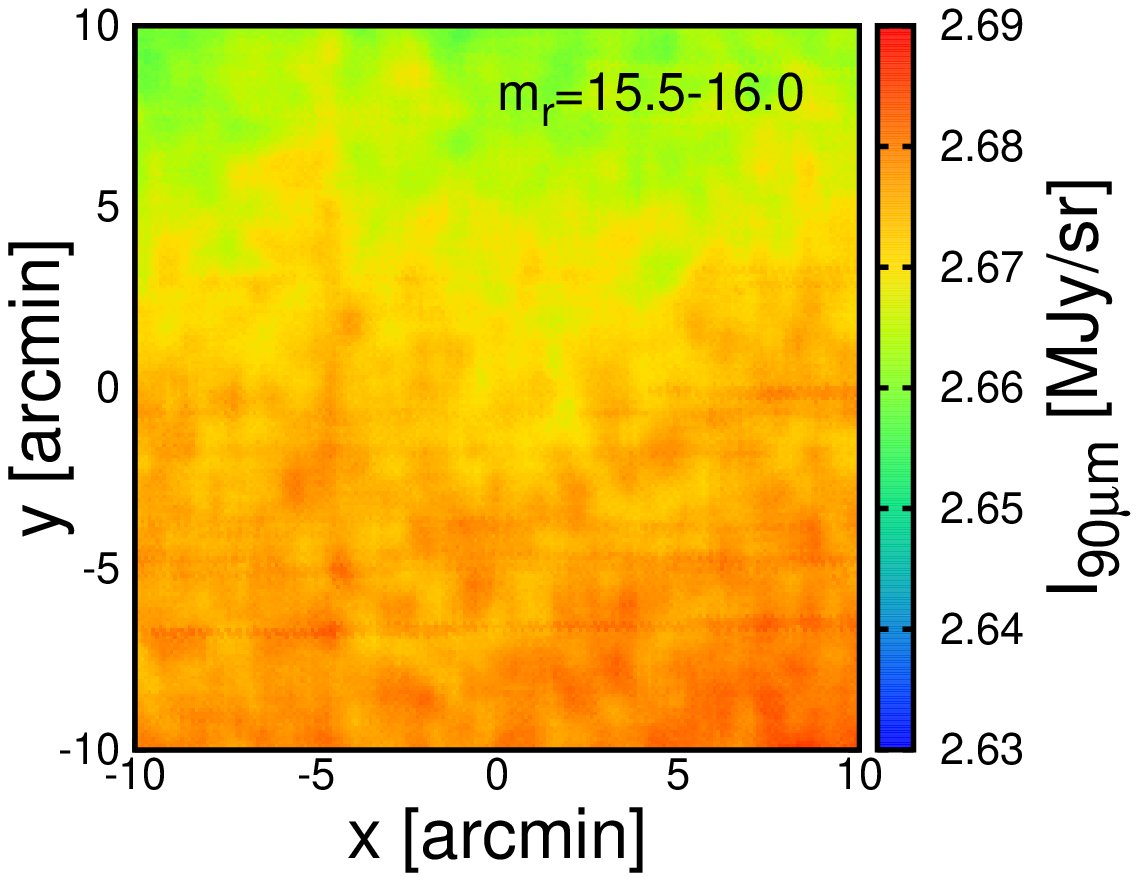}
    \FigureFile(55mm,55mm){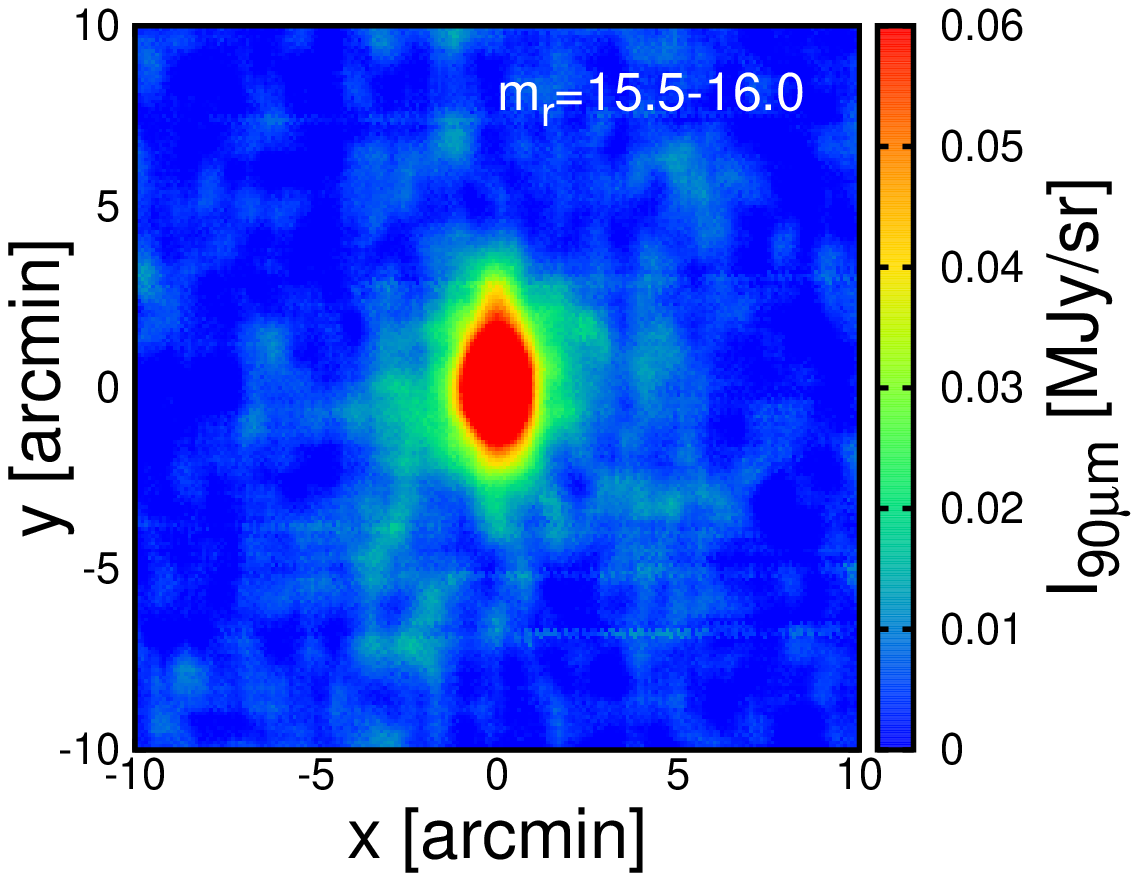}
    \FigureFile(55mm,55mm){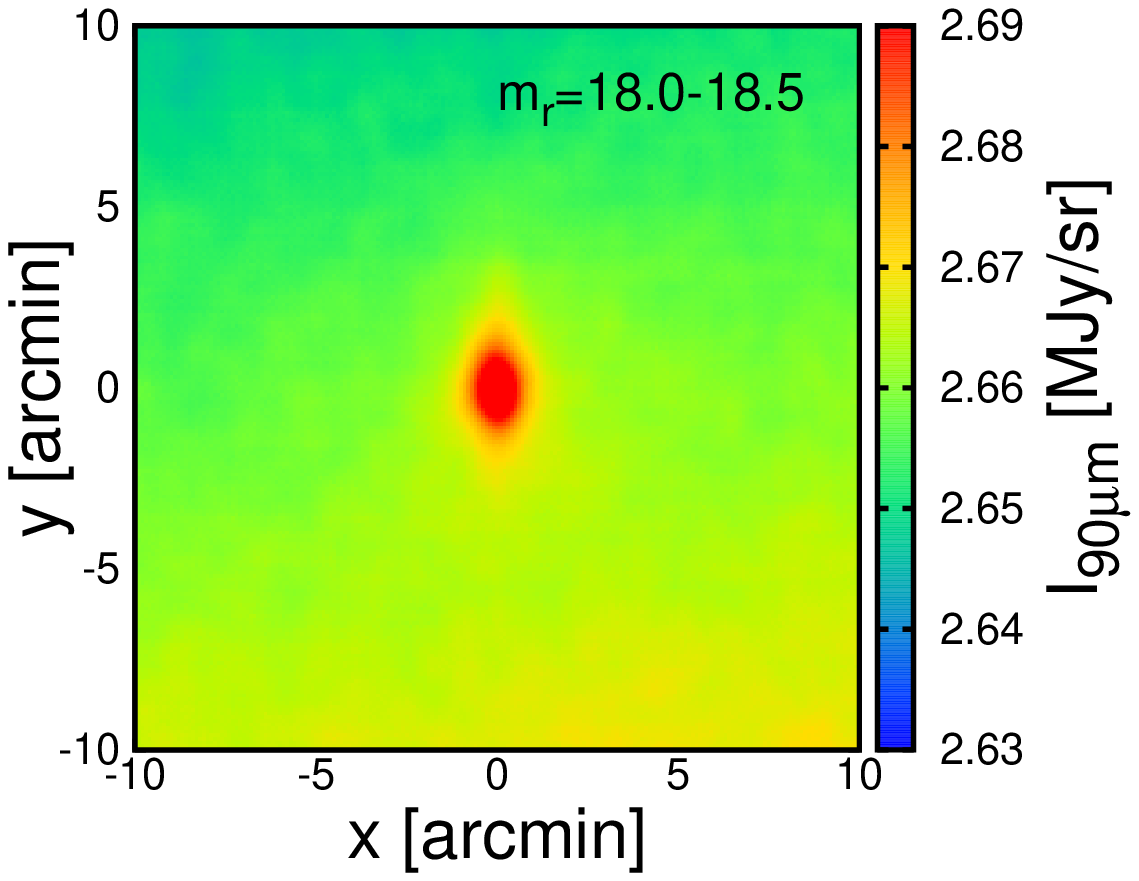}
    \FigureFile(55mm,55mm){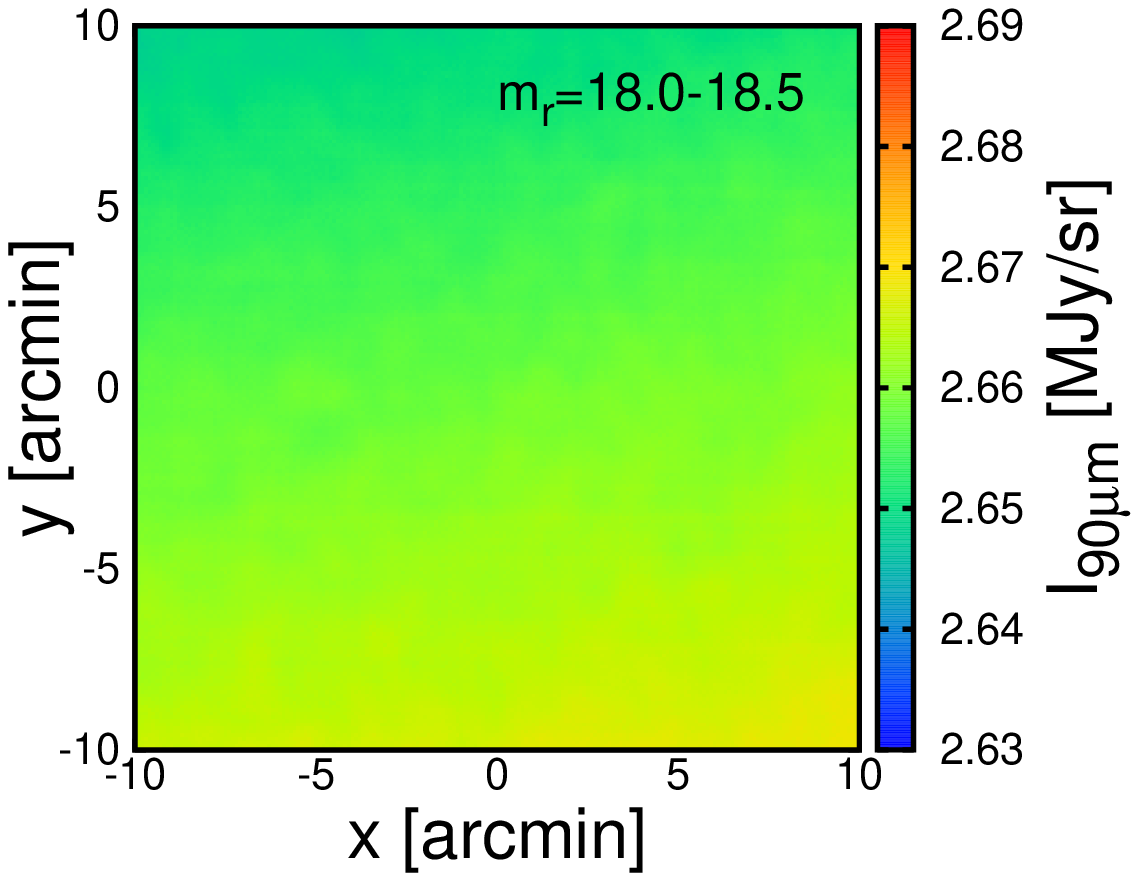}
    \FigureFile(55mm,55mm){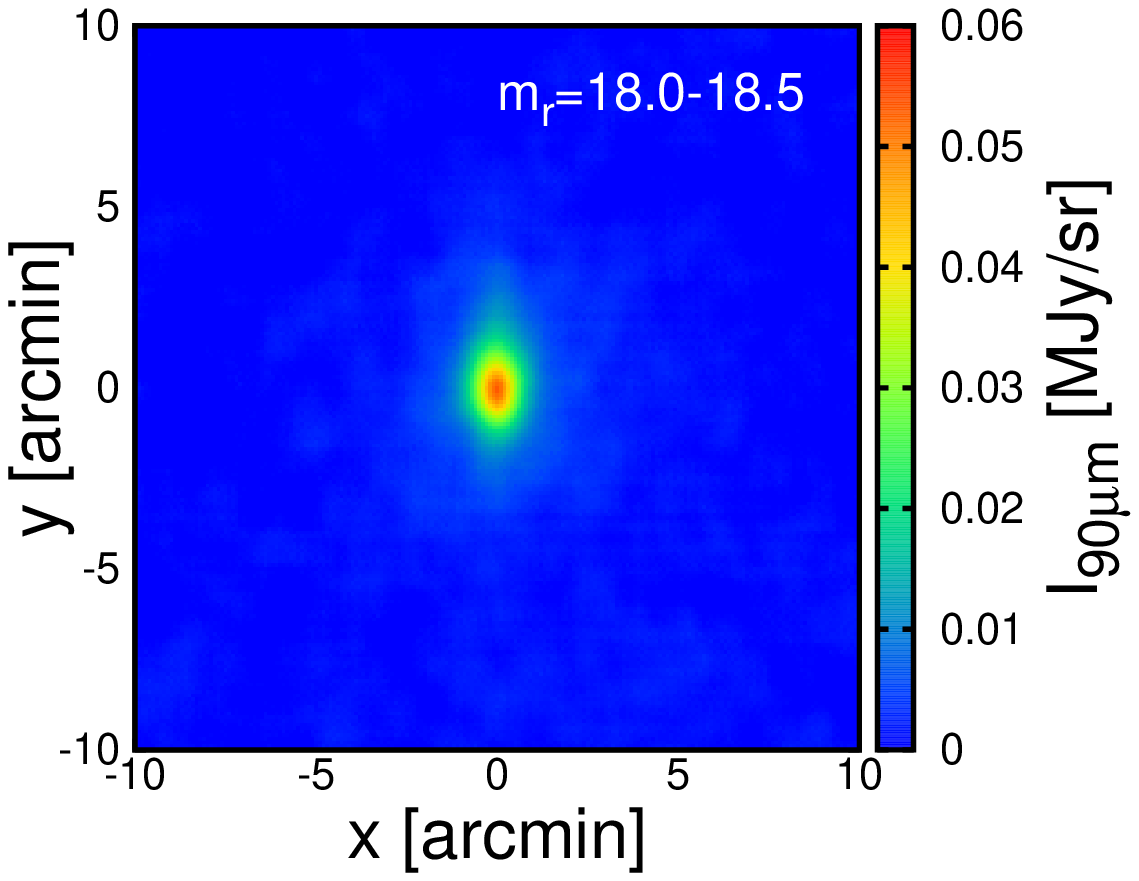}
    \FigureFile(55mm,55mm){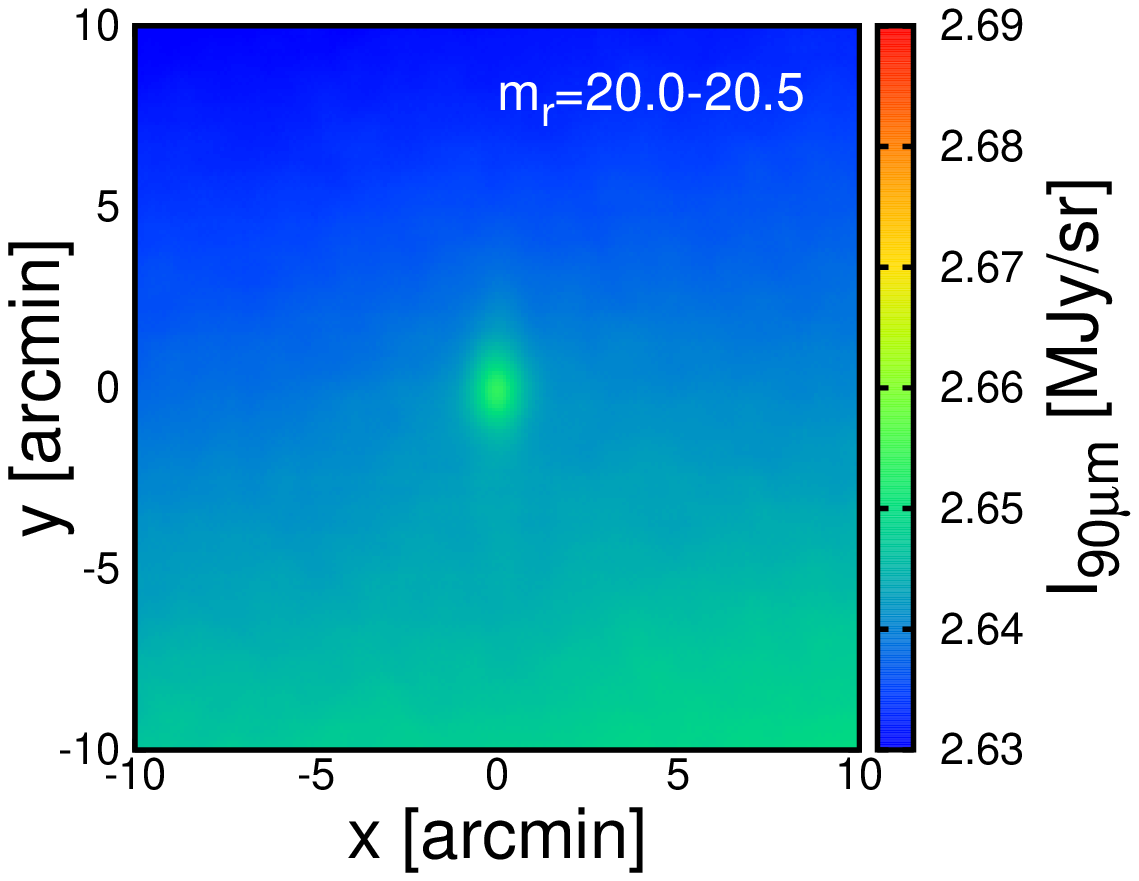}
    \FigureFile(55mm,55mm){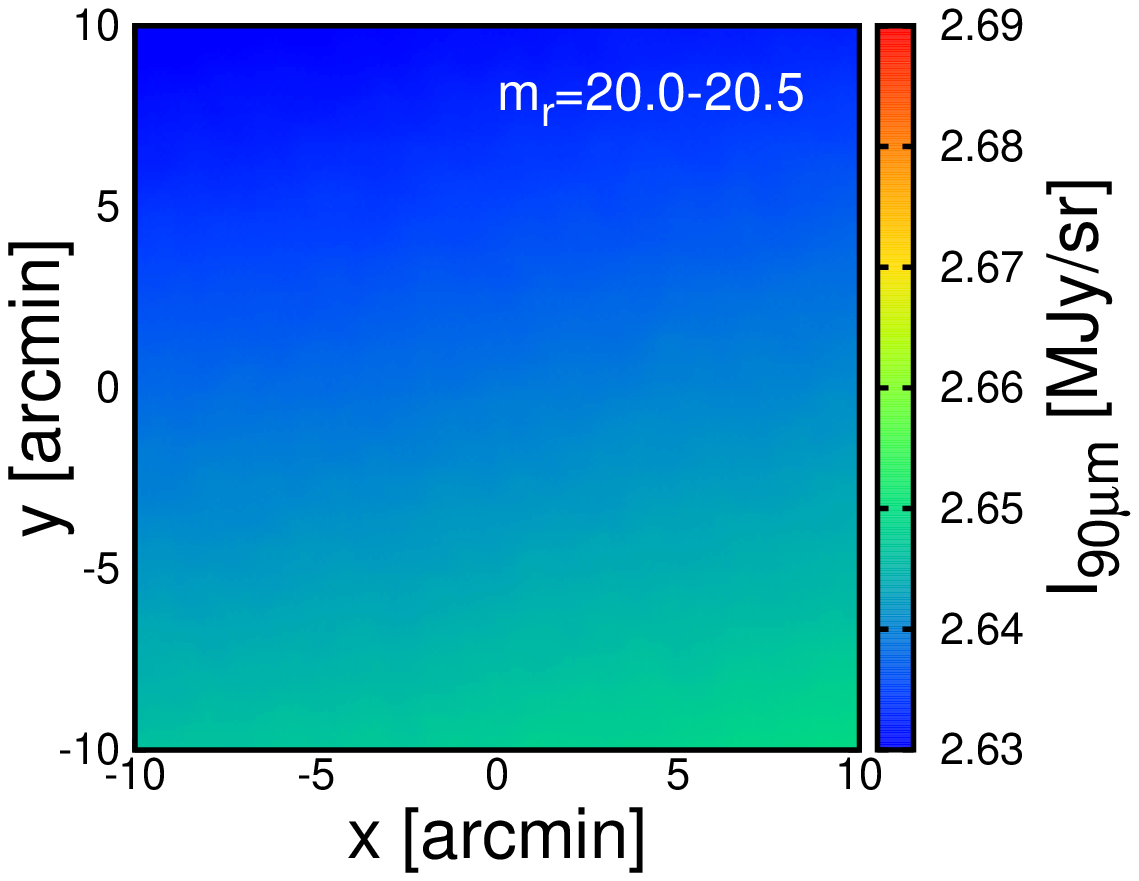}
    \FigureFile(55mm,55mm){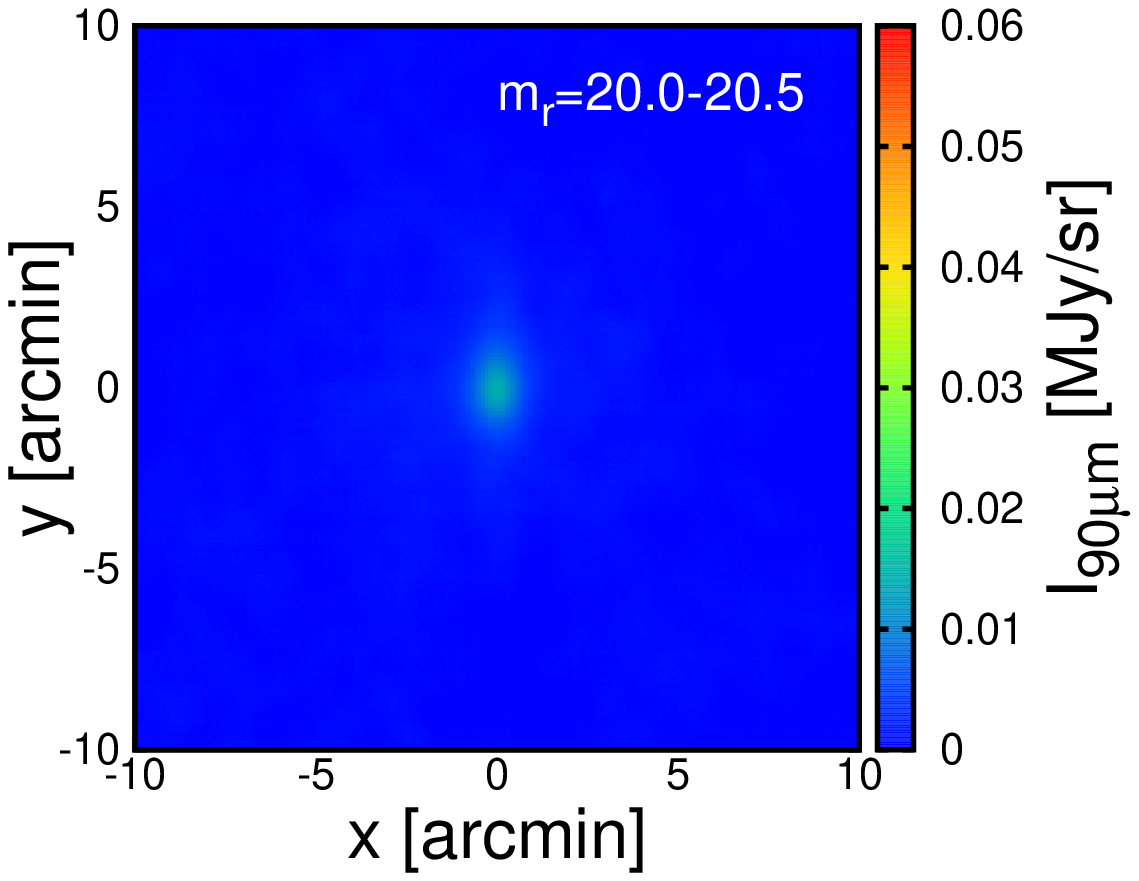}
\end{center}
\caption{Left, middle and right panels correspond to the raw
  stacked images, foreground templates, and the stacked images
  subtracting the foreground templates, respectively. The upper,
  center, and lower panels represent the result for $m_r=15.5$-$16.0$,
  $18.0$-$18.5$, and $20.0$-$20.5$, respectively. }
\label{fig:galaxy-background}
\end{figure*}

\subsection{Stacking Method\label{sec:stack}}

Since the detection limits of SDSS are much deeper than those of FIS,
the majority of the SDSS galaxies cannot be individually resolved in the
FIS maps.  The stacking analysis on the FIS maps, however, enables us to
statistically measure the average FIR emission of the SDSS galaxies.
Our method for the stacking analysis is basically the same as that
adopted by KYS13, except for that we mask several bright objects and
subtract the zodiacal and Galactic foreground as will be described
below.

Consider first the 90$\mum$ data.  We divide the entire sample of SDSS
galaxies according to their $r$-band magnitude and stack $20'\times20'$
images of the FIS map centered at the positions of SDSS galaxies in each
magnitude bin.  In this procedure, we evaluate the value of $I_{90\mum}$
on $6''\times6''$ pixels over the $20'\times20'$ images, adopting a
cloud-in-cell linear interpolation of the four nearest neighbors in the
${15''}\times{15''}$ pixels of the original FIS data.  We first stack
those images fixing the $y$-direction to the ecliptic longitudes so as
to retain the anisotropic PSF of the FIS.  Since the FIS map does not
remove point sources unlike the SFD data, we exclude those images that
contain pixels with $I_{90\mum}>200{\rm MJy/sr}$ to avoid the
contamination due to bright point sources.  We confirmed that our main
result does not change even when we adopt $I_{90\mum}>100{\rm MJy/sr}$
or $I_{90\mum}>1000{\rm MJy/sr}$.  Our criterion excludes several large
nearby galaxies, and effectively masks about 2 ${\rm deg}^2$ in total.
Table \ref{table:numbers-of-galaxies} lists the number of SDSS galaxies
labelled as ``GALAXY'', selected as our sample, and stacked after
removing the regions with bright sources.

Figure \ref{fig:galaxy-background} illustrates our stacking procedure in
three different magnitude bins as an example. The left panels indicate
the raw stacked results before circular averaging. In addition to the
elongated source images, there is a systematic large-scale gradient
along the $y$-axis, which originate from Galactic dust, and residual
zodiacal light.

In order to remove the gradient component, we shift the centers of the
images by $+{20''}$ and $-{20''}$ along the $x$-axis, {\it i.e.,}
constant ecliptic latitude, from the source galaxy position, and repeat
the same stacking. The average of those off-source images (central
panels) is used as templates of the Galactic and zodiacal foreground,
and is subtracted from the corresponding raw stacked images (left
panels).

The right panels of figure \ref{fig:galaxy-background} show the stacked
images after the subtraction.  Clearly the large-scale gradient is
successfully removed, and the signal of SDSS galaxies becomes more
pronounced; note the different range of $I_{90\mum}$ shown in the right
color scales.  All the analysis below is based on the stacked images
corresponding to the right panels of figure \ref{fig:galaxy-background}.

\begin{figure*}[t]
\begin{center}
    \FigureFile(55mm,55mm){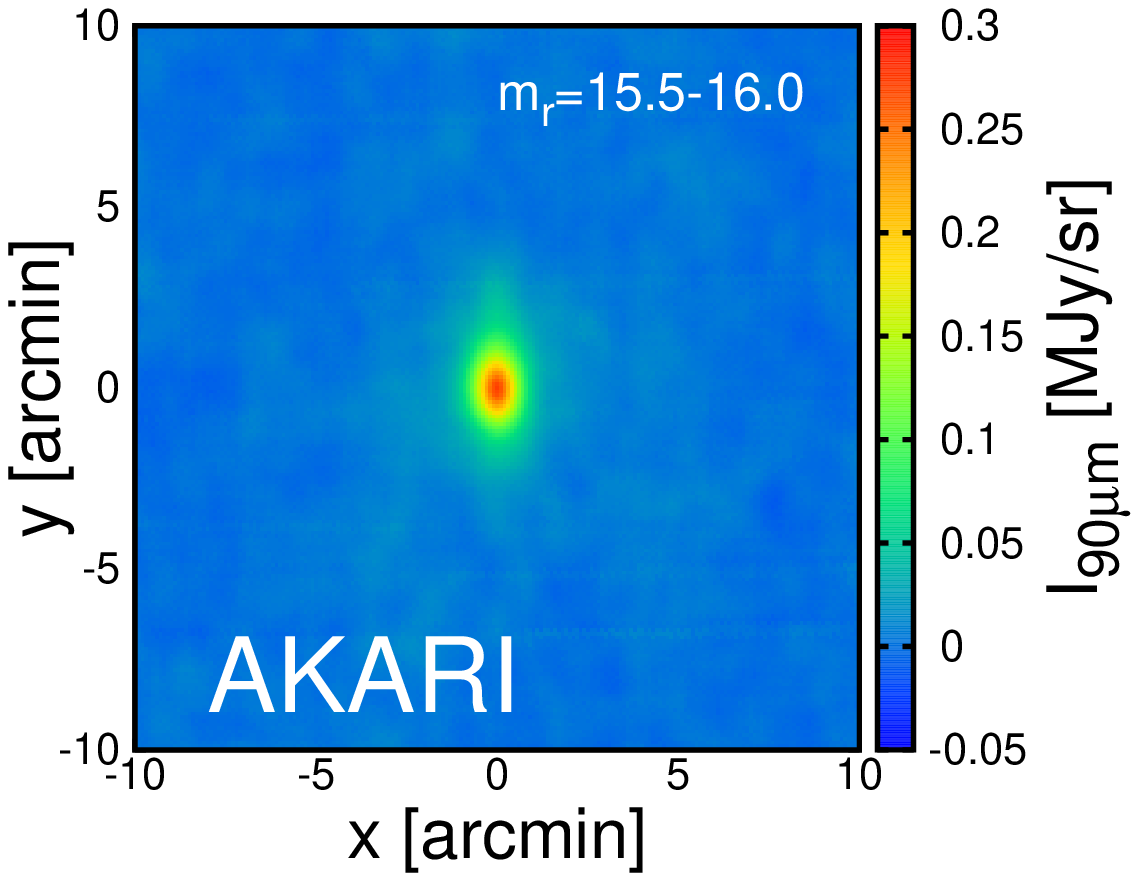}
    \FigureFile(55mm,55mm){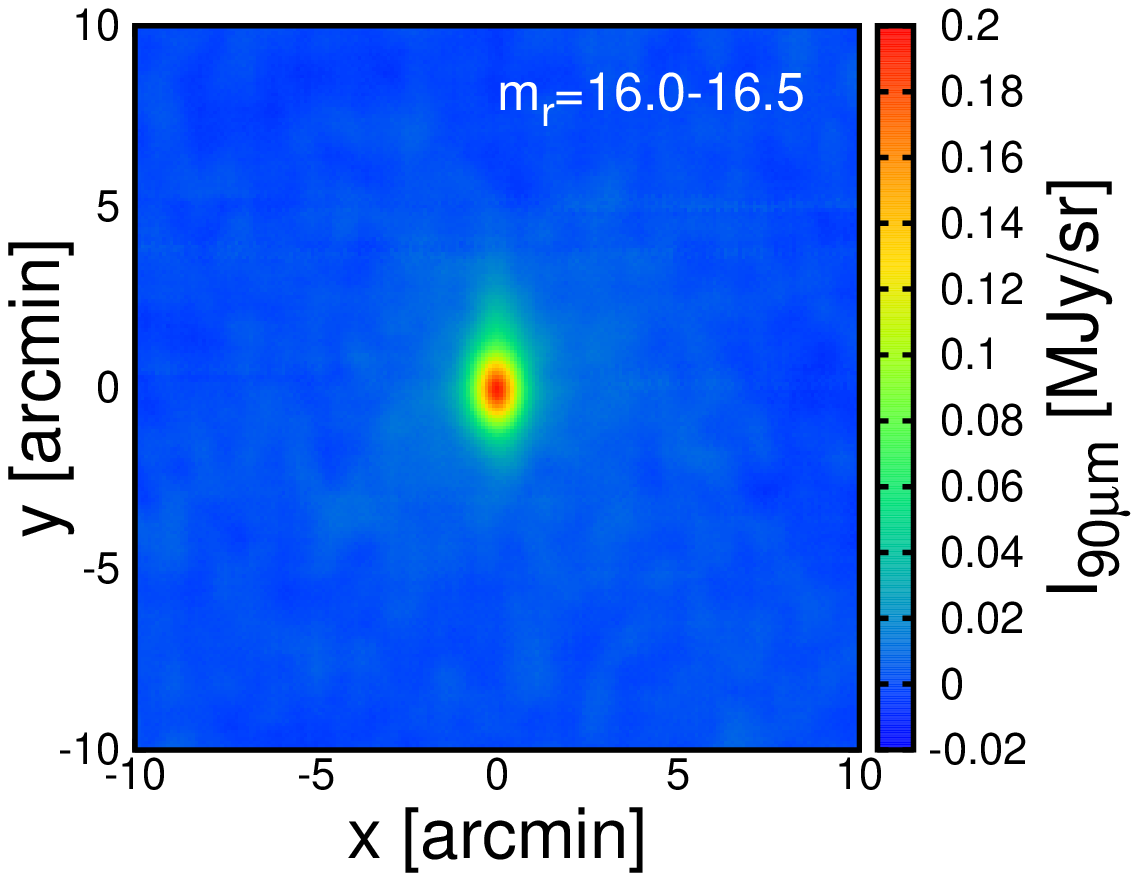}
    \FigureFile(55mm,55mm){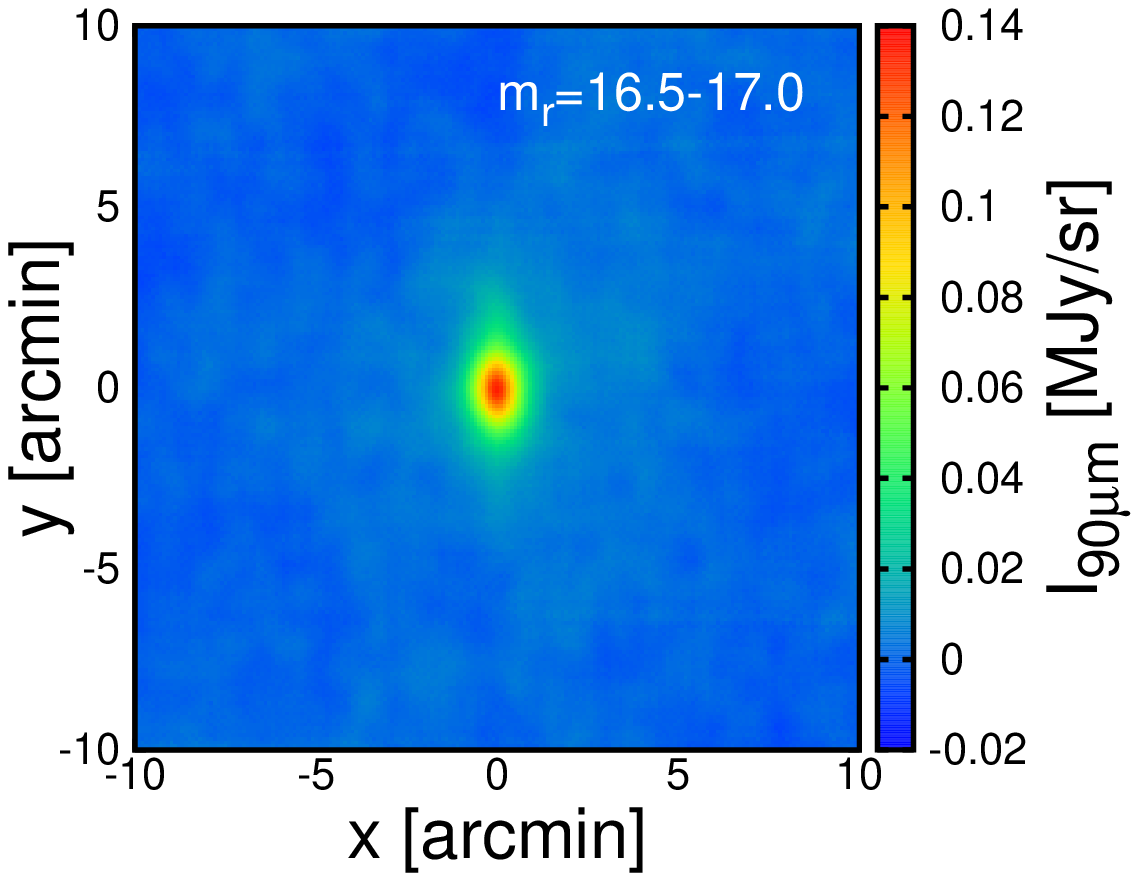}
    \FigureFile(55mm,55mm){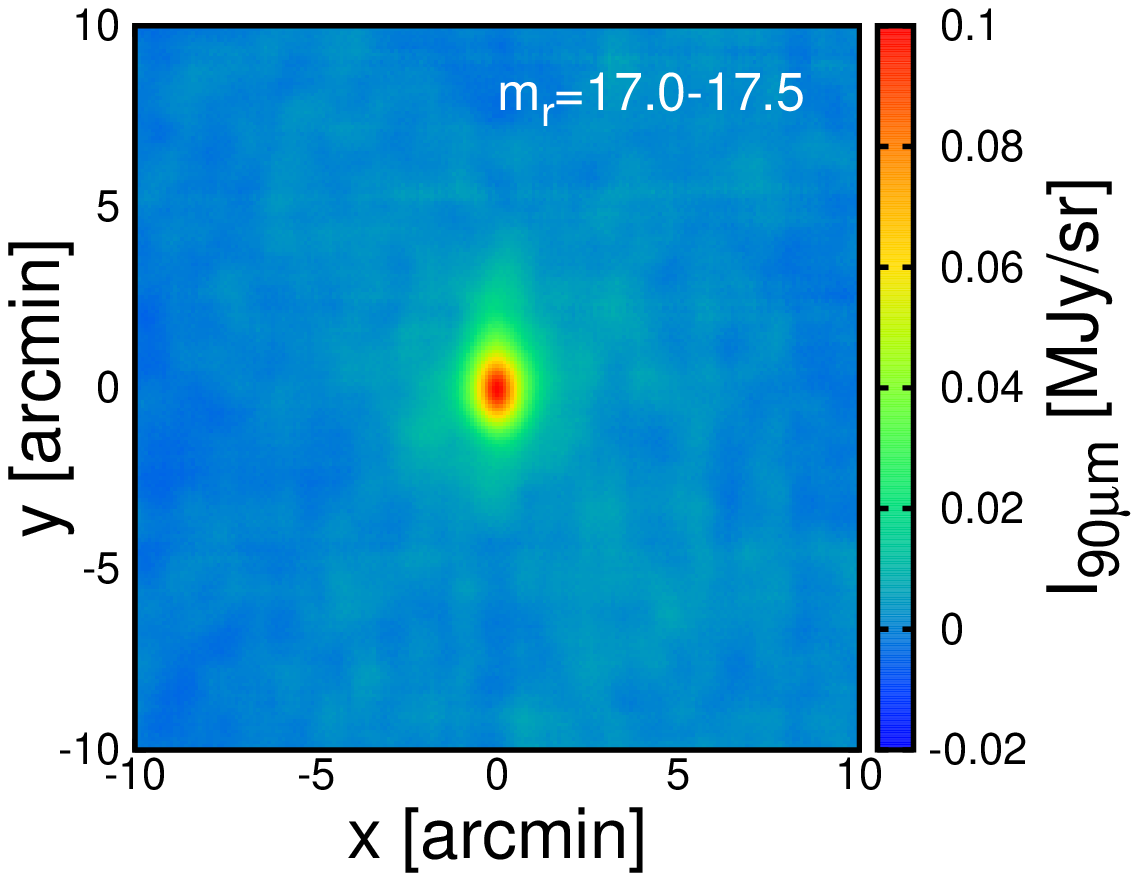}
    \FigureFile(55mm,55mm){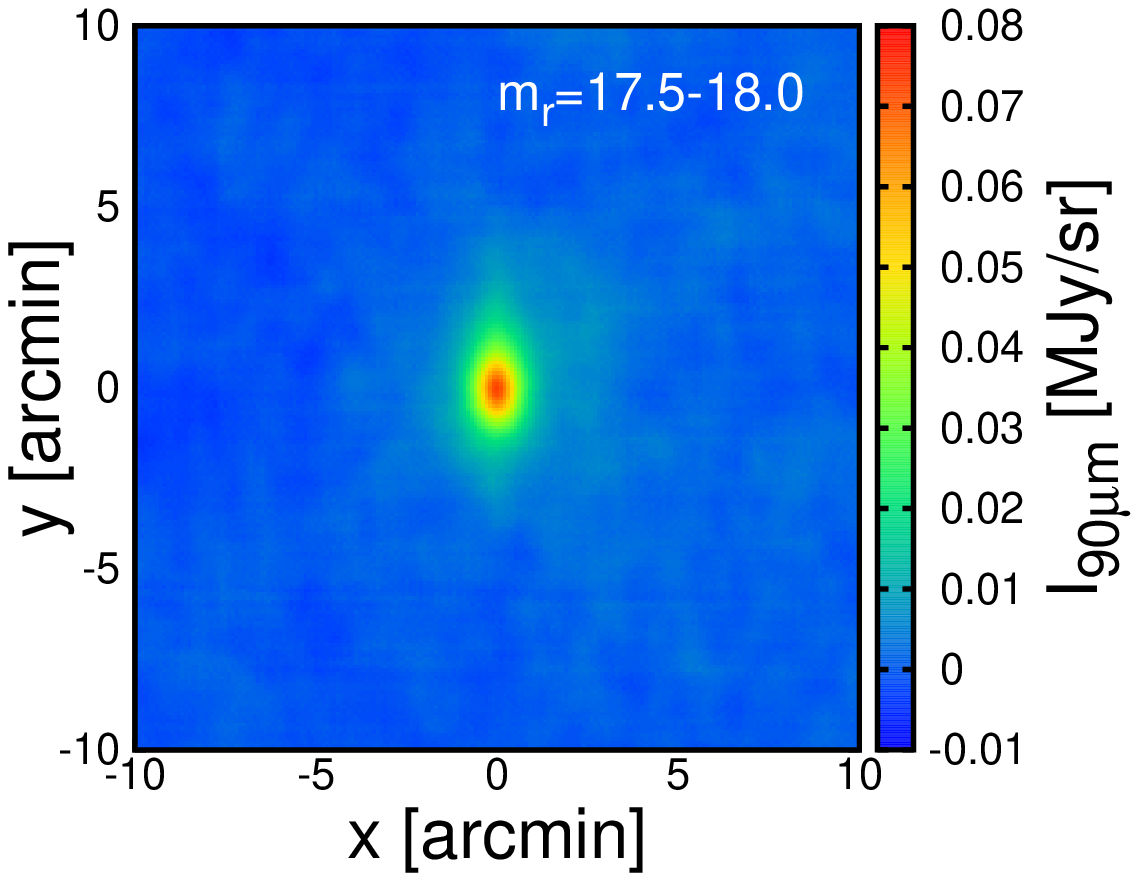}
    \FigureFile(55mm,55mm){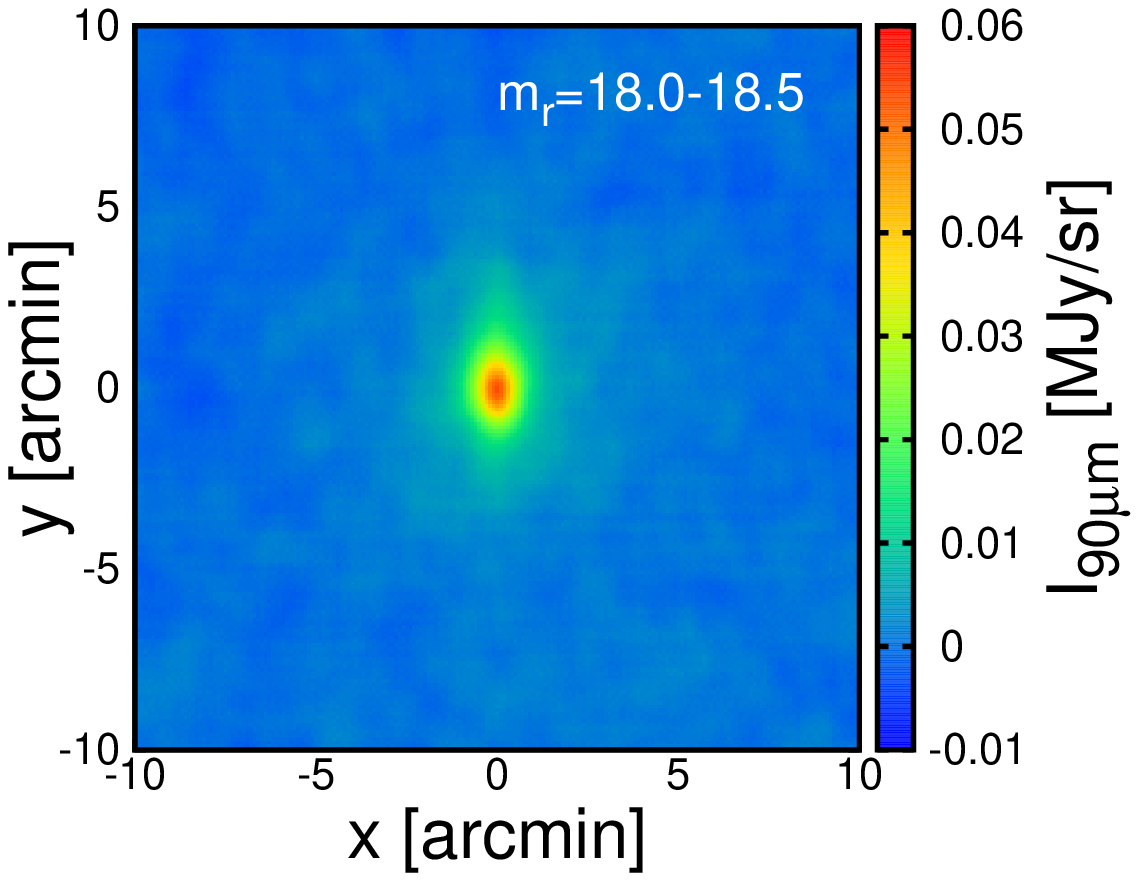}
    \FigureFile(55mm,55mm){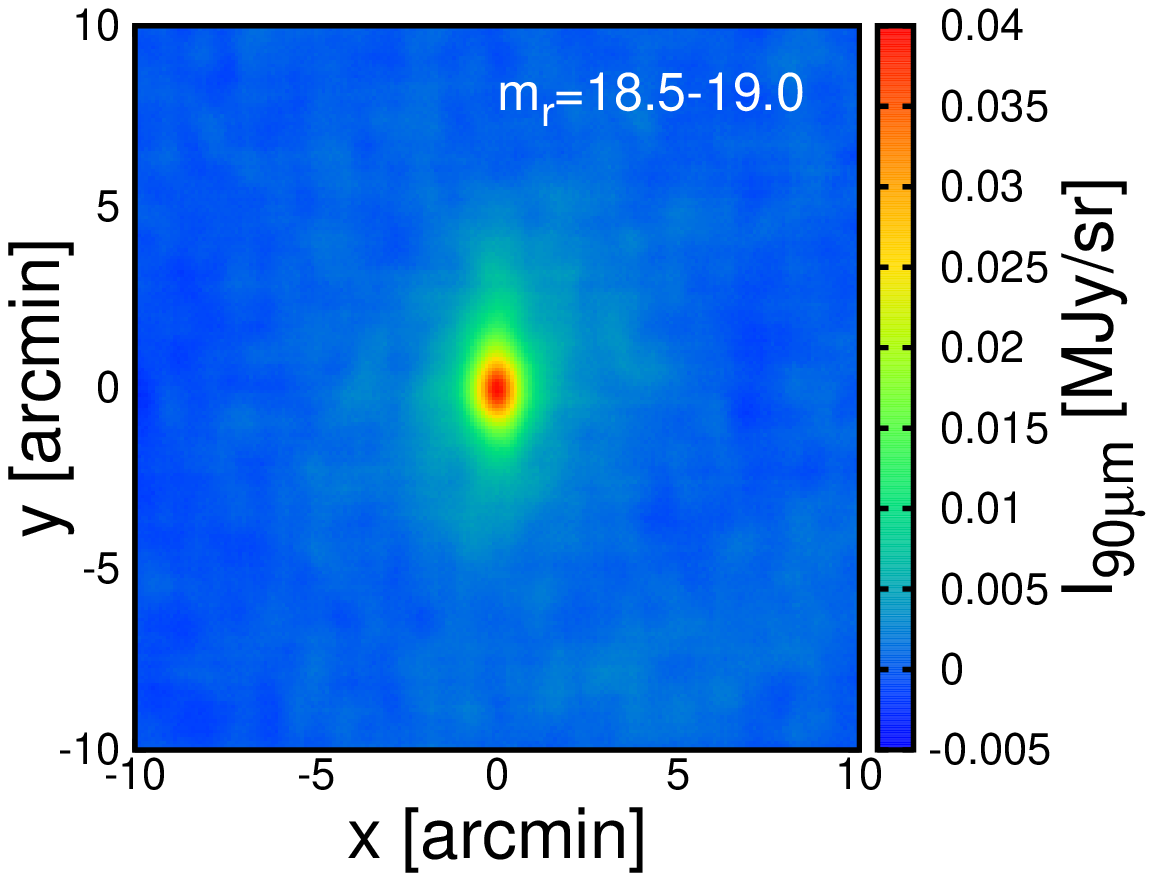}
    \FigureFile(55mm,55mm){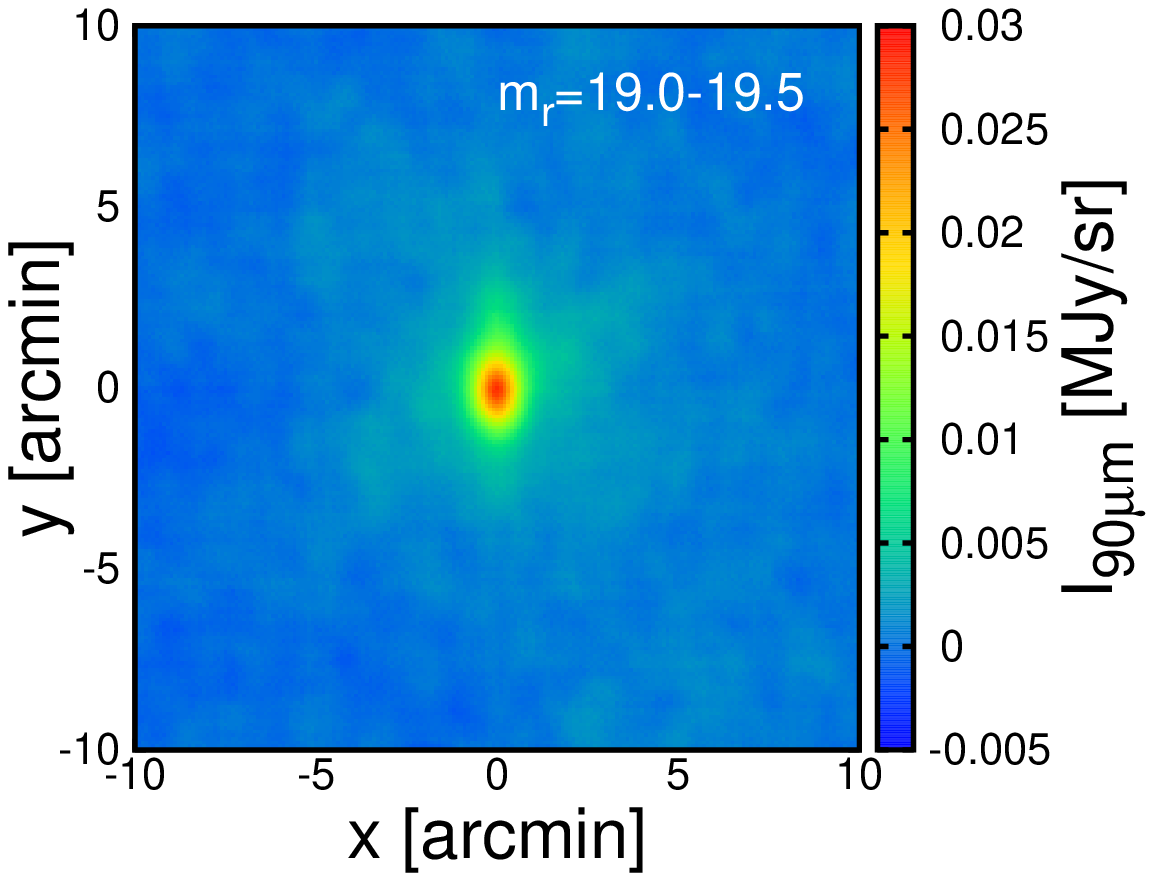}
    \FigureFile(55mm,55mm){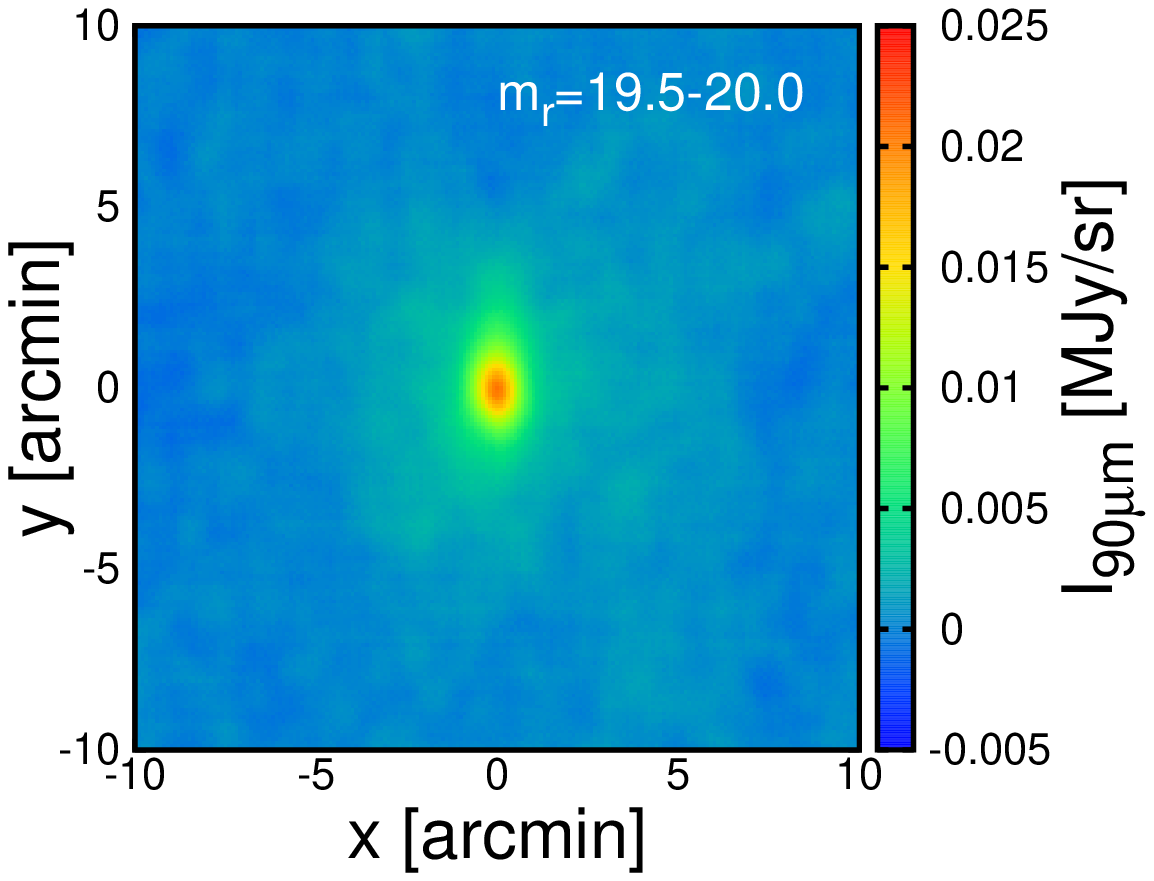}
    \FigureFile(55mm,55mm){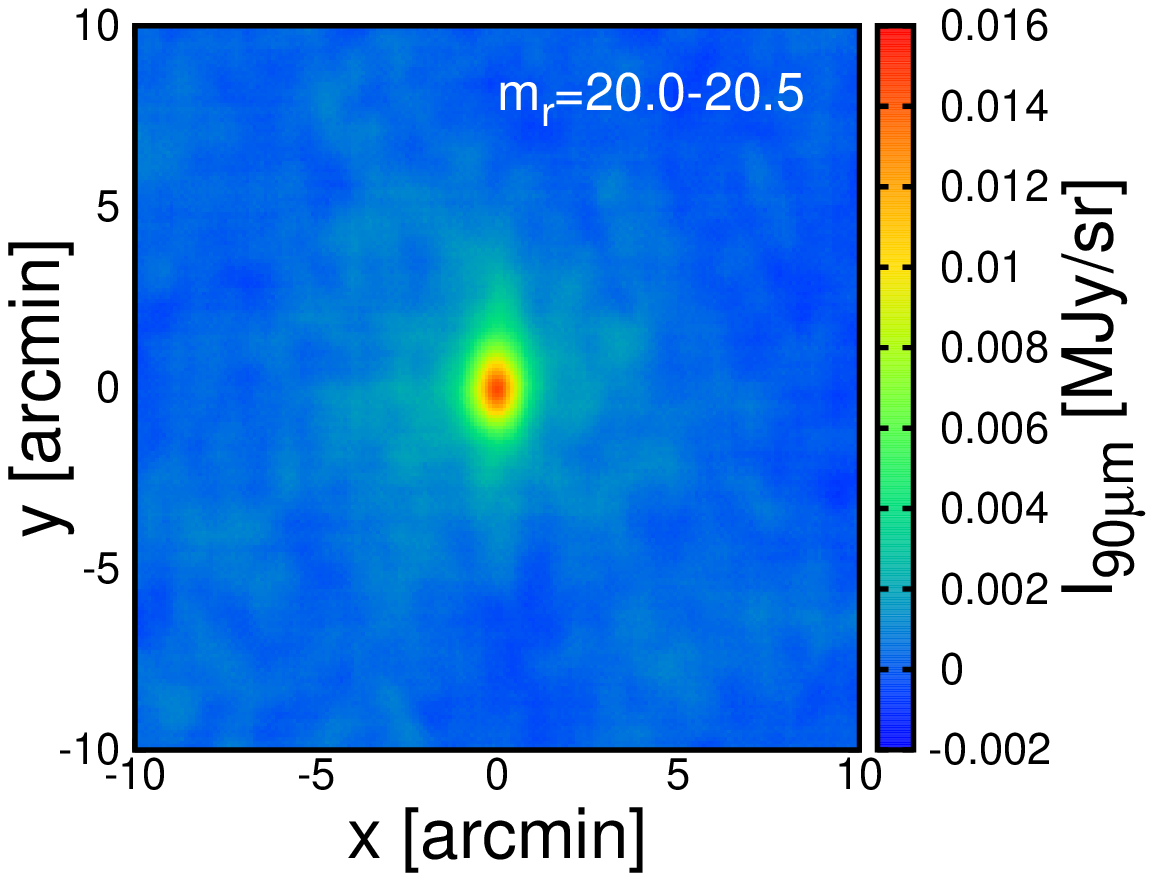}
    \FigureFile(55mm,55mm){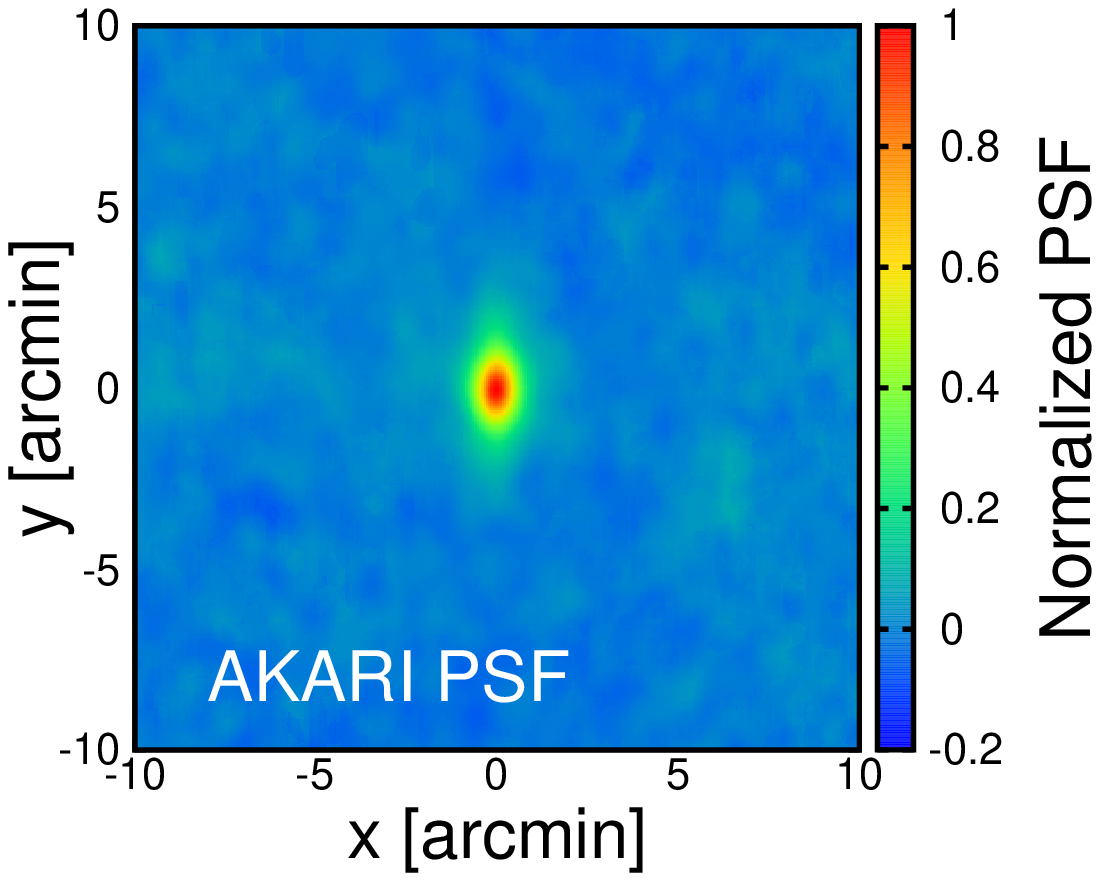}
    \FigureFile(55mm,55mm){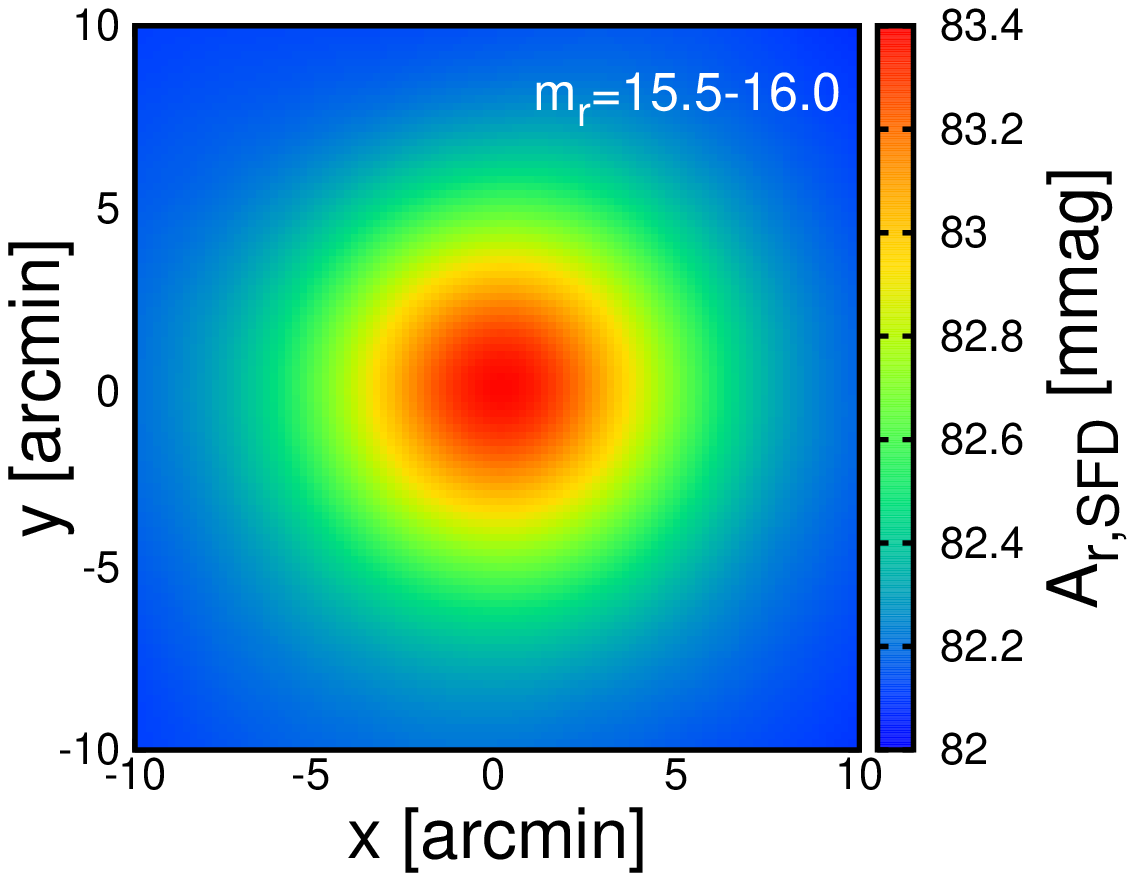}
\end{center}
\caption{Stacked images of the AKARI FIS map for $20'\times20'$ centered at
  SDSS galaxies of different $r$-band magnitudes ($m_r=15.5$-$20.5$)
  in 0.5 magnitude bin.  The bottom center panel correspond the PSF of
  FIS at 90$\mum$ by \citet{Arimatsu2014}.  For comparison, the bottom
  right panel shows the stacked images of SFD extinction map by
  KYS13.}
\label{fig:galaxy-diff-mag}
\end{figure*}

\section{Decomposing Stacked Images \label{sec:results}}

Figure \ref{fig:galaxy-diff-mag} displays the stacked images of the SDSS
galaxies according to the method described in \S\ref{sec:stack} for ten
different $r$-band magnitudes (Table \ref{table:numbers-of-galaxies}).
For reference, we show the stacked image of bright stars (bottom middle
panel) that should correspond to the PSF of FIS, and the stacked image
on IRAS (bottom right panel) as well for comparison.  The resulting
stacked images exhibit prominent signatures of emission associated with
the SDSS galaxies located at the origin, and indicate clearly that the
angular resolution of FIS is much better than that of IRAS.

When a typical galaxy of a radius 10kpc is located at the median
redshift of the SDSS sample ($\langle z\rangle=0.36$,
\citealt{Dodelson2002}), the angular size is about ${2''}$, much
smaller than the size of PSFs of FIS.  Thus we neglect the intrinsic
size of galaxies, which is justified even visually from the comparison
between the stacked galaxy and star images (figure
\ref{fig:galaxy-diff-mag}).

\begin{figure*}[bt]
\begin{center}
    \FigureFile(80mm,80mm){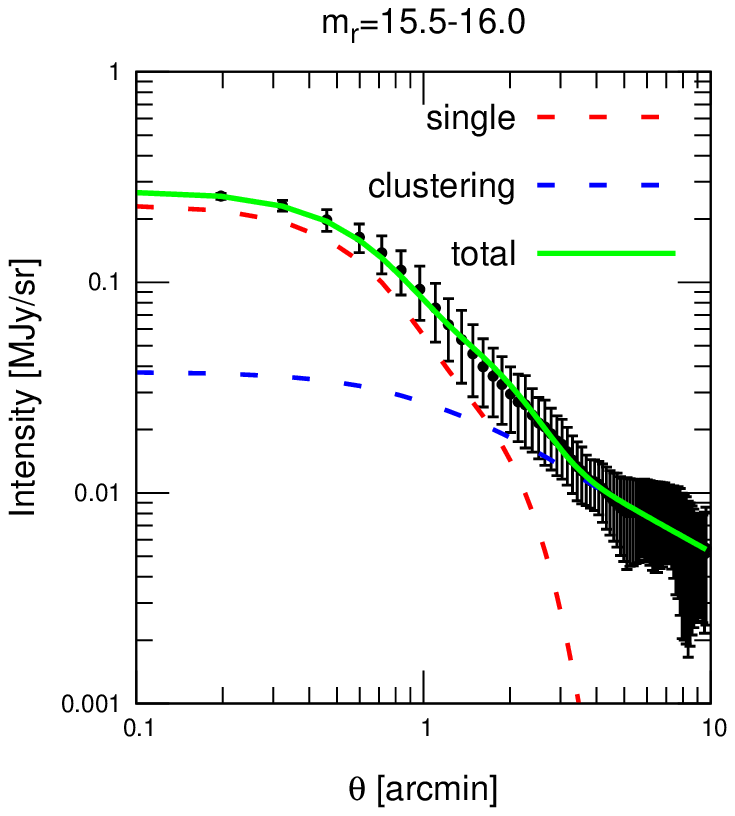}
    \FigureFile(80mm,80mm){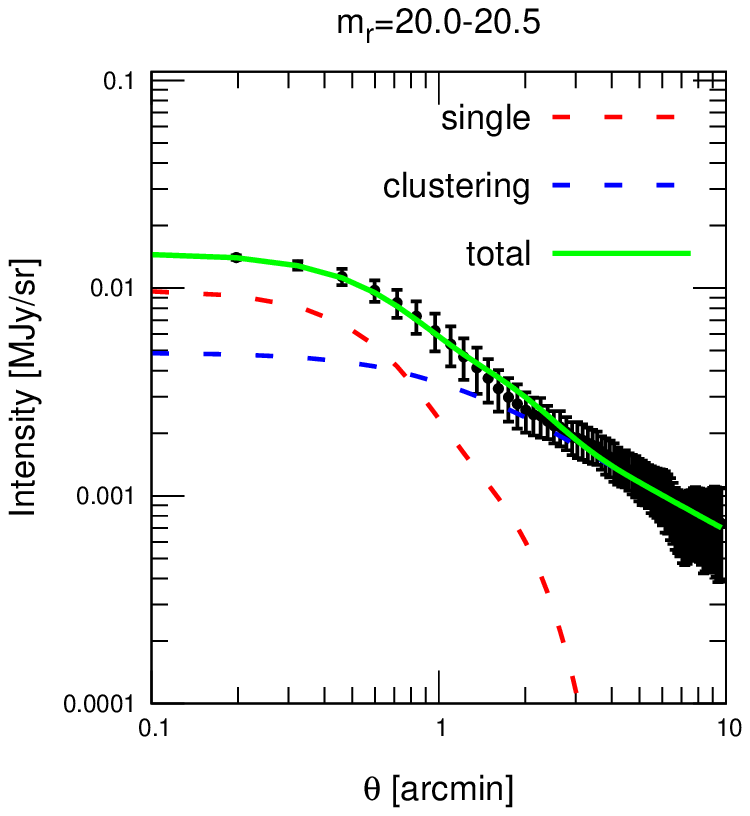}
\end{center} 
\caption{Circular-averaged radial profiles of the stacked images of SDSS
  galaxies corresponding to figure \ref{fig:galaxy-diff-mag}. Green
  solid, red, and blue dashed curves indicate the best-fit model of
  equations (\ref{eq:average-galaxy-profile}),
  (\ref{eq:single-galaxy-profile}), and
  (\ref{eq:clustering-galaxy-profile2}), respectively.  The quoted
  error-bars indicate the rms within each radial bin.}
  \label{fig:profile}
\end{figure*}

To quantitatively characterize the FIR emission of SDSS galaxies, we
compute the circular-averaged radial profiles of the stacked images,
$\Sigma_{\rm g}^{\rm tot}({\boldsymbol \theta};m_r)$, in Figure
\ref{fig:profile}; $15.5<m_r<16.0$ (left panel) and $20.0<m_r<20.5$
(right panel).

The quoted error-bars represent the rms within each radial bin, $\Delta
\theta = 7''.76$, and are dominated by the anisotropy of the PSF.
Following KYS13, we then model the measured radial profiles as the sum
of three components:
\begin{eqnarray}
\label{eq:average-galaxy-profile}
\Sigma_{\rm g}^{\rm tot}({\boldsymbol \theta};m_r) =
\Sigma_{\rm g}^{\rm s}({\boldsymbol \theta};m_r) 
+\Sigma_{\rm g}^{\rm c}({\boldsymbol \theta};m_r)
+ \Delta C(m_r),
\end{eqnarray}
where $\Sigma_{\rm g}^{\rm s}({\boldsymbol \theta};m_r)$ is the single
term (the contribution from central galaxies in the stacked images),
$\Sigma_{\rm g}^{\rm c}({\boldsymbol \theta};m_r)$ is the clustering
term (the contribution from clustered galaxies around the central
galaxies), and $\Delta C$ is the residual offset level of the average
foreground emission after subtracting the foreground templates.  The
specific expression for the single and clustering term will be given in
equation (\ref{eq:single-galaxy-profile}) and
(\ref{eq:clustering-galaxy-profile2}), respectively, and the residual
offset level will be discussed in detail later this section.

Since we neglect the intrinsic size of galaxies, the single term is
represented by the PSF profile, equation (\ref{eq:double-PSF-profile}):
\begin{eqnarray}
\label{eq:single-galaxy-profile}
\Sigma_{\rm g}^{\rm s}({\boldsymbol \theta};m_r)  =
\Sigma_{\rm g}^{\rm s0}(m_r)W_2(\theta).
\end{eqnarray}
Then the clustering term is written as:
\begin{eqnarray}
\label{eq:clustering-galaxy-profile}
\Sigma_{\rm g}^{\rm c}({\boldsymbol \theta};m_r) =
\iint dm' d{\boldsymbol \varphi}
~\Sigma_{\rm g}^{\rm s}({\boldsymbol \theta-\boldsymbol \varphi};m') 
w_{\rm g}({\boldsymbol \varphi};m',m_r)
n_{\rm g}(m') ,
\end{eqnarray}
where $w_{\rm g}({\boldsymbol \varphi};m',m_r)$ is the angular
correlation function of galaxies with magnitude $m'$ located at
${\boldsymbol \varphi}$ from the central galaxy with magnitude $m_r$,
and $n_{\rm g}(m')\equiv dN_{\rm g}(m')/dm'$ is the differential galaxy
number density.

\begin{figure*}[h]
\begin{center}
    \FigureFile(100mm,100mm){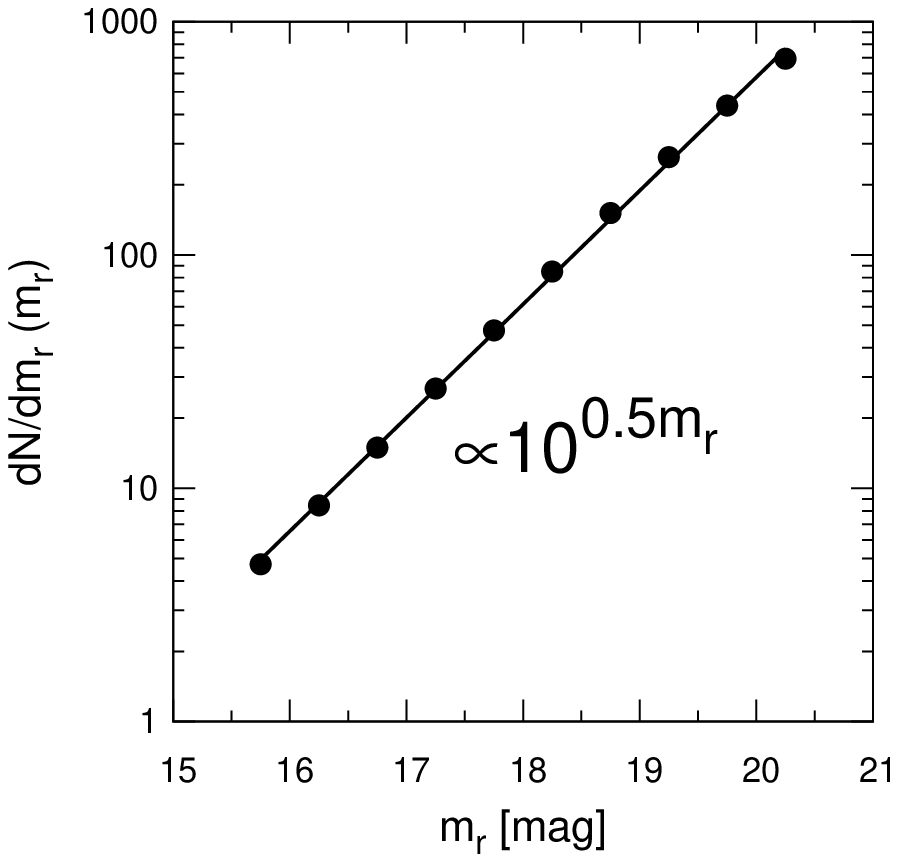}
 \end{center} 
\caption{Differential number count of the SDSS galaxies as a
  functional of $r$-band magnitudes. The symbols and solid line show
  the SDSS data and their best-fit, respectively.}
\label{fig:dndm}
\begin{center}
    \FigureFile(100mm,100mm){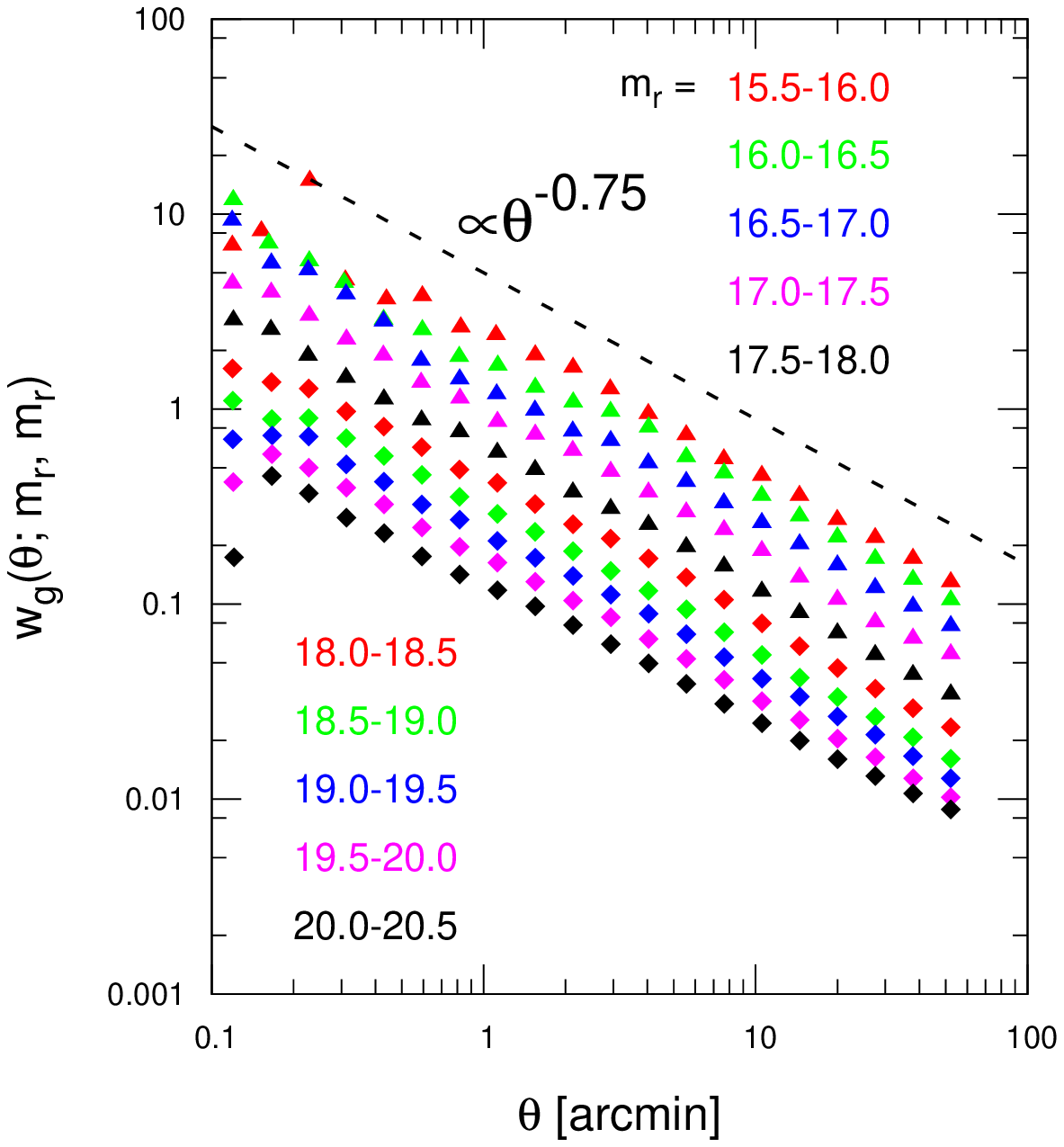}
 \end{center} 
\caption{The angular-correlation function of SDSS galaxies for each
  magnitude bin.  The black dashed line indicates $\gamma=0.75$.}
\label{fig:mainACF}
\end{figure*}

These two functions are directly measured from the SDSS galaxies; figure
\ref{fig:dndm} plots the differential galaxy number density, and figure
\ref{fig:mainACF} shows the angular-correlation function of SDSS
galaxies for $m'=m_r$.  We also find that the angular-correlation
function for $m'\neq m_r$ obeys the same power-law, and can be
approximated as
\begin{eqnarray}
\label{eq:wg}
w_{\rm g}({\boldsymbol\varphi};m',m_r)
=K(m',m_r) \left(\frac{\varphi}{\varphi_0}\right)^{-\gamma},
\end{eqnarray}
with $\gamma$ and $\varphi_0$ being independent of $m_r$.  We adopt
$\gamma=0.75$, which is confirmed to be valid for $\varphi<1^{\circ}$
(\citealt{Connolly2002}; \citealt{Scranton2002}).  For reference, the
dashed line in figure \ref{fig:mainACF} shows $\gamma=0.75$.

Substituting equations (\ref{eq:single-galaxy-profile}) and
(\ref{eq:wg}) into equation (\ref{eq:clustering-galaxy-profile}) 
yields
\begin{eqnarray}
\label{eq:clustering-galaxy-profile2}
\Sigma_{\rm g}^{\rm c}({\boldsymbol \theta};m_r) =
\Sigma_{\rm g}^{\rm c0}(m_r)W^{\rm c}(\theta), 
\end{eqnarray}
\begin{eqnarray}
\label{eq:sigmac0}
\Sigma_{\rm g}^{\rm c0}(m_r) =
2\pi\left(A\sigma_1^{2-\gamma}+(1-A)\sigma_2^{2-\gamma}\right)
\left(\frac{\varphi_0}{\sqrt{2}}\right)^{\gamma}
\Gamma\left(1-\frac{\gamma}{2}\right)
\int dm' \Sigma_{\rm g}^{\rm s0}(m')
K(m',m_r)n_{\rm g}(m'),
\end{eqnarray}
\begin{eqnarray}
\label{eq:Wc}
W^{\rm c}(\theta) &\equiv& B\exp\left(-\frac{\theta^2}{2\sigma_1^2}\right)
{}_1F_1\left(1-\frac{\gamma}{2};1;\frac{\theta^2}{2\sigma_1^2}\right)+\left(1-B\right)\exp\left(-\frac{\theta^2}{2\sigma_2^2}\right)
{}_1F_1\left(1-\frac{\gamma}{2};1;\frac{\theta^2}{2\sigma_2^2}\right),
\end{eqnarray}
where $_1F_1\left(\alpha;\beta;x\right)$ is a confluent hypergeometric
function, and
\begin{eqnarray}
B\equiv\frac{A\sigma_1^{2-\gamma}}
{A\sigma_1^{2-\gamma}+(1-A)\sigma_2^{2-\gamma}}.
\end{eqnarray}

In this section, we do not use equation (\ref{eq:sigmac0}), and fit
equations (\ref{eq:average-galaxy-profile}),
(\ref{eq:single-galaxy-profile}), (\ref{eq:clustering-galaxy-profile2}),
and (\ref{eq:Wc}) to the observed circular-averaged radial profiles of
the stacked images separately for each magnitude by varying the three
parameters $\Sigma_{\rm g}^{\rm s0}(m_r)$, $\Sigma_{\rm g}^{\rm
s0}(m_r)$, and $\Delta C(m_r)$. The comparison of the resulting
$\Sigma_{\rm g}^{\rm c0}(m_r)$ with equation (\ref{eq:sigmac0}) will be
considered in the next section.

\clearpage

\begin{figure}[bt]
\begin{center}
   \FigureFile(100mm,100mm){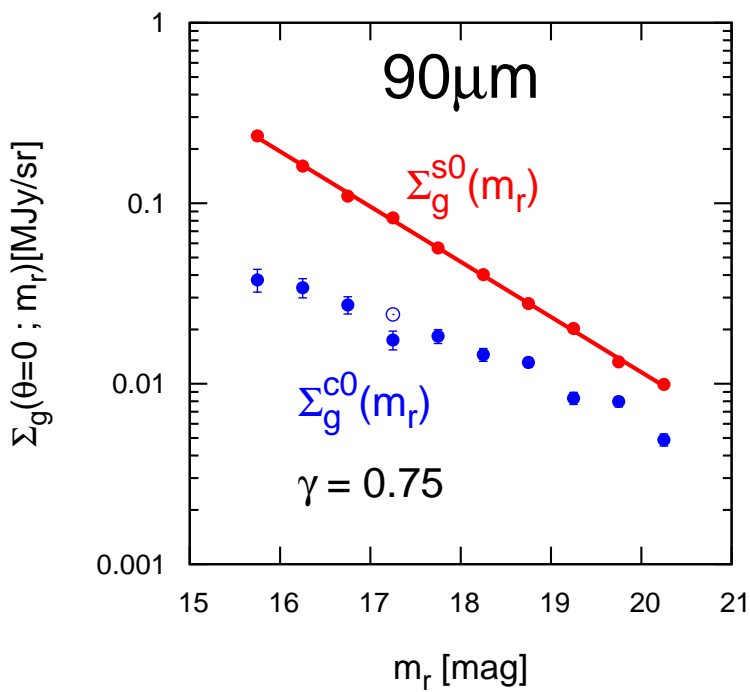}
\end{center} 
\caption{Best-fit parameters characterizing the FIR emission of
  galaxies against their $r$-band magnitude. The quoted error-bars are
  computed from 1000 subsamples of jackknife resampling.  Red and blue
  solid symbols indicate the best-fit values of $\Sigma_{\rm g}^{\rm
    s0}(m_r)$, and $\Sigma_{\rm g}^{\rm c0}(m_r)$, respectively,
  assuming $\gamma=0.75$.  The red solid line indicates to the
  best-fit line of the best-fit value of $\Sigma_{\rm g}^{\rm
    s0}(m_r)$ corresponding to the equation (\ref{eq:sigmas0model}).}
 \label{fig:bestfit}
\end{figure}

\begin{figure}[bt]
\begin{center}
   \FigureFile(100mm,100mm){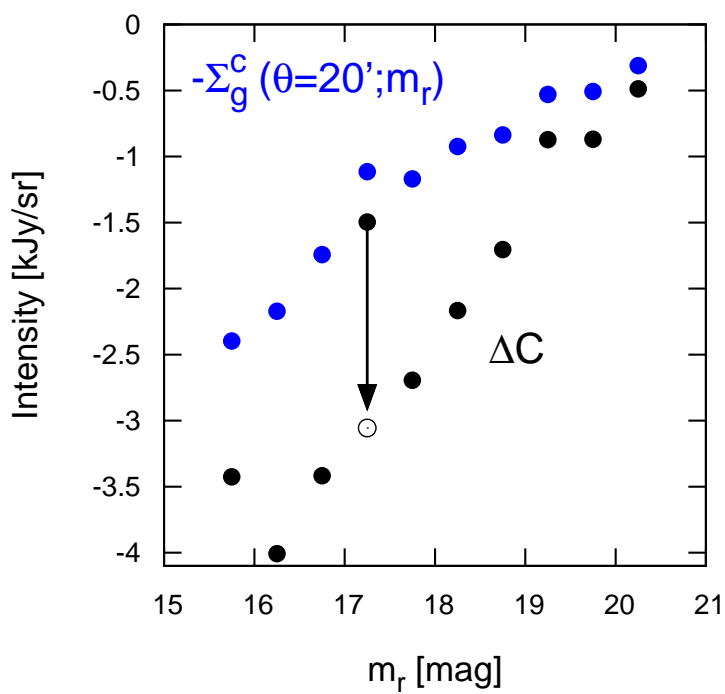}
\end{center}
\caption{Best-fit parameters of residual constant term
  against their $r$-band magnitude (black solid symbols)
  and negative value of clustering term at $\theta=20'$, -$\Sigma_{\rm g}^{\rm c}(\theta=\timeform{20'};m_r)$
  (blue solid symbols).}
 \label{fig:bestfit-foreground}
\end{figure}
\clearpage

The resulting best-fits are plotted in green solid, red dashed, and blue
dashed curves, respectively, in figure \ref{fig:profile}.  Figure
\ref{fig:bestfit} shows the best-fits of $\Sigma_{\rm g}^{\rm s0}(m_r)$
and $\Sigma_{\rm g}^{\rm c0}(m_r)$ against $m_r$, where the quoted
error-bars are computed from 1000 subsamples of the jackknife resampling
method.  The results indicate that the single term dominates the
clustering term around the center of the stacked images.  This is in
marked contrast to the result of KYS13 for IRAS at 100$\mum$; their
figures 8, 9 and 10 indicate that $\Sigma_{\rm g}^{\rm c0}(m_r)$ is
significantly larger than $\Sigma_{\rm g}^{\rm s0}(m_r)$. Thanks to the
better angular resolution of AKARI, we are able to separate the
prominent emission from the central galaxy from the surrounding diffuse
component due to the nearby galaxies.  This also implies that the AKARI
data enable us to measure the contribution of single term more robustly,
in a less model dependent fashion.  The difference between AKARI and
IRAS will be discussed in detail later in this section.

Figure \ref{fig:bestfit-foreground} also indicates that $\Delta C$ at
$17.0 < m_r < 17.5$ bin significant departs from the systematic trend of
the entire samples.  A similar gap is also seen for $\Sigma_{\rm g}^{\rm
c0}(m_r)$ in figure \ref{fig:bestfit}.  Indeed, if we use the value of
$\Delta C$ at $17.0 < m_r < 17.5$ bin from the interpolation of the
adjacent bins (black open circle in figure
\ref{fig:bestfit-foreground}), the resulting best-fit of $\Sigma_{\rm
g}^{\rm c0}(m_r)$ also matches the other trend (blue open circle in
figure \ref{fig:bestfit}).  While we are not yet able to identify the
reason of this behavior, we do not correct for this in the analysis below.

Figure \ref{fig:bestfit-foreground} shows that the best-fit values of
$\Delta C$ systematically increase against $m_r$.  Such $m_r$-dependence
of $\Delta C$ seems unphysical, and is likely to result from our
foreground (Galactic component) subtraction procedure.  Our foreground
templates are computed from the stacked images at $\pm 20'$ away from
the center, and thus the radial profiles of the foreground-subtracted
data should vanish at $\theta = 20'$ even though the clustering term
would still extend beyond the scale.  As a consequence of the
over-correction, we thought that the residual of the foreground $\Delta
C$ would be equal to $-\Sigma_{\rm g}^{\rm c}(\theta=20';m_r)$.  As
shown in figure \ref{fig:bestfit-foreground}, however, this does not
hold exactly even though the qualitative trend is consistent; the blue
and black filled circles differ by a factor of two.  We suspect that
this difference should come from the difficulty to decompose the weak
radial dependence of the clustering term from the constant offset due to
the Galactic dust in our model fit. In any case, we made sure that this
dependence does not affect our main conclusion, and we do not consider
it in what follows.

\begin{figure*}[h]
\begin{center}
    \FigureFile(60mm,60mm){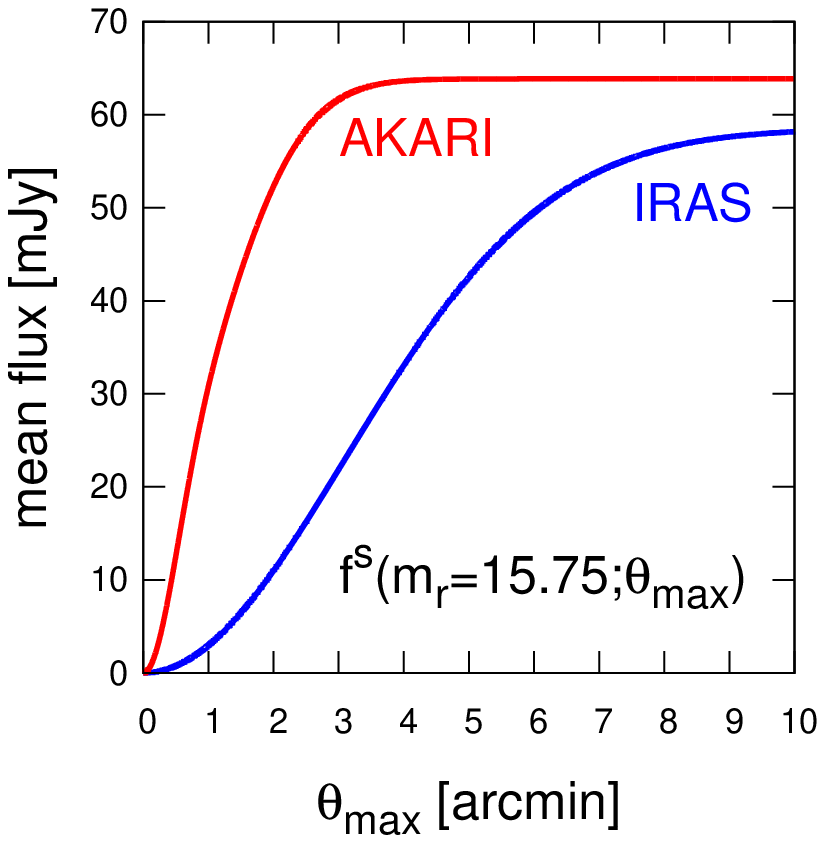}
    \FigureFile(60mm,60mm){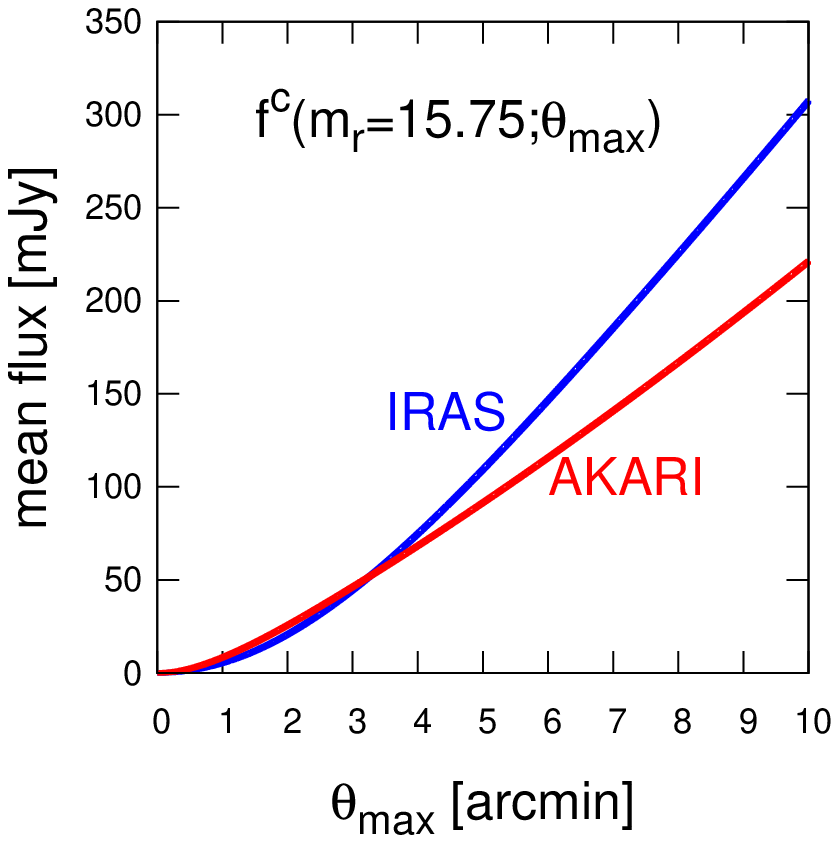}
 \end{center}
\caption{Flux of galaxies integrated within $\theta_{\rm max}$ for AKARI
(red) and IRAS (blue).  Left and right panels correspond to the flux of
single term and clustering term, respectively.}  \label{fig:convergion}
\end{figure*}

It is also interesting to compare the present result with that obtained
by KYS13 for IRAS.  For that purpose, we compute the fluxes of the
single galaxy at the center of the stacked map:
\begin{eqnarray}
\label{eq:flux-single}
f^s(m_r; \theta_{\rm max}) &=& 
\int \Sigma_{\rm g}^{\rm s}({\boldsymbol \theta};m_r)d{\boldsymbol \theta} 
\simeq 2\pi\int_0^{\theta_{\rm max}} 
\Sigma_{\rm g}^{\rm s0}(m_r)W_2(\theta)\theta d\theta, 
\end{eqnarray}
and the contribution of neighbor galaxies around the center galaxy:
\begin{eqnarray}
\label{eq:flux-clustering}
f^c(m_r; \theta_{\rm max}) &=& 
\int \Sigma_{\rm g}^{\rm c}({\boldsymbol \theta};m_r)d{\boldsymbol \theta} 
\simeq 2\pi\int_0^{\theta_{\rm max}} 
\Sigma_{\rm g}^{\rm c0}(m_r)W^{\rm c}(\theta)\theta d\theta. 
\end{eqnarray}
The integral (\ref{eq:flux-clustering}) does not converge for
$w(\theta)\propto \theta^{-0.75}$, but the power-law is not valid for
$\theta > 1^{\circ}$ in any way.  Thus we introduce an upper limit,
$\theta_{\rm max}$, in the integral.  Its value is somewhat arbitrary,
but we adopt $\theta_{\rm max}={10'}$ because we only use the profiles
of stacked images to $\theta={10'}$ for the fitting.
Figure \ref{fig:convergion} shows the flux as a function of $\theta_{\rm max}$ in equations (\ref{eq:flux-single}) and (\ref{eq:flux-clustering}).
The left and right panels correspond to $f^{\rm s}(m_r; \theta_{\rm max})$ and $f^{\rm c}(m_r; \theta_{\rm max})$, respectively, 
for AKARI (red line) and IRAS (blue line).
Figure \ref{fig:convergion} suggests that the fluxes of the single term for both AKARI and IRAS
converge at $\theta_{\rm max}\sim\timeform{10'}$,
while the fluxes of the clustering term diverge.

\clearpage

\begin{figure*}[bt]
\begin{center}
    \FigureFile(100mm,100mm){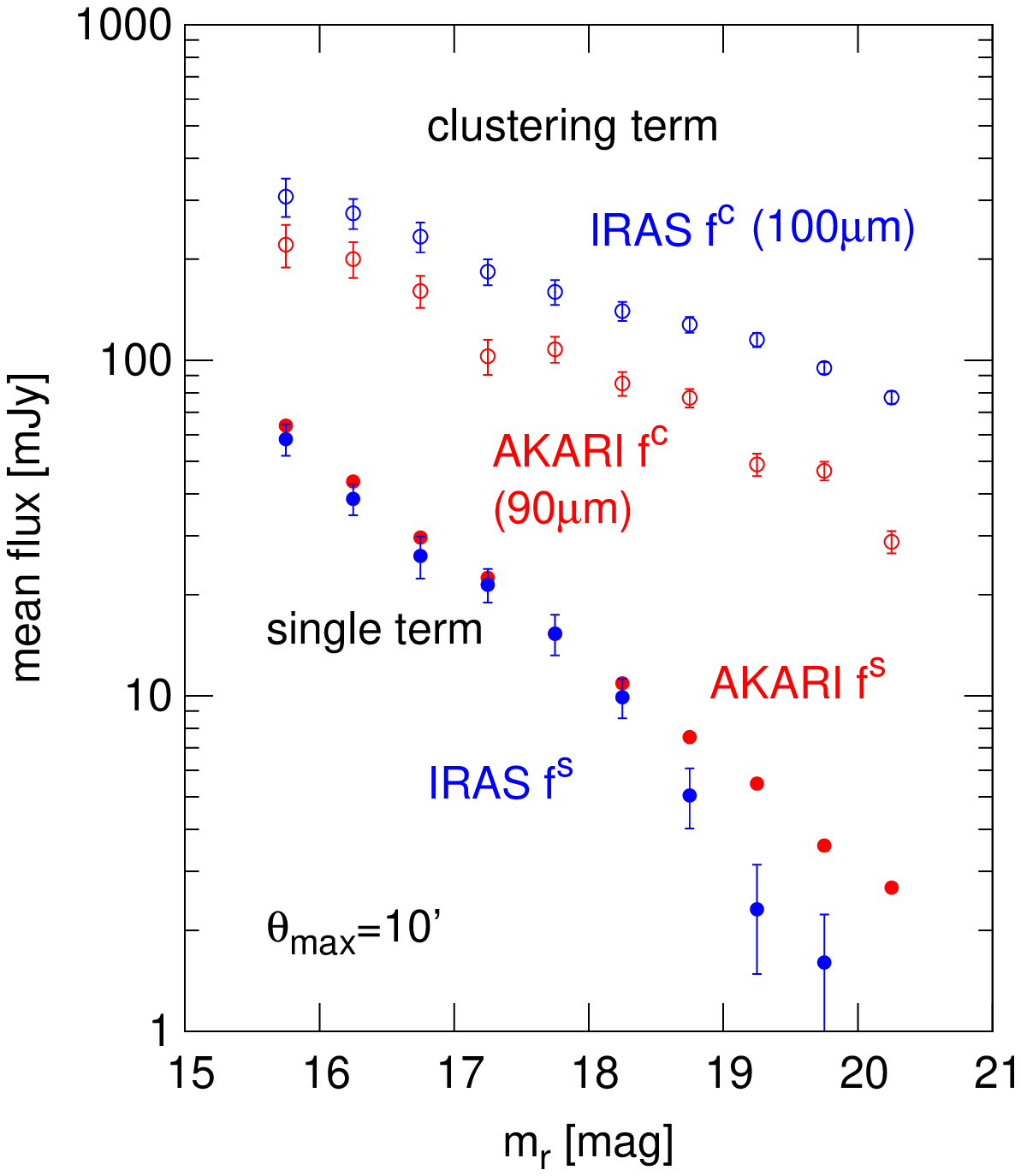}
 \end{center}
\caption{The flux of galaxies integrated up to $\theta_{\rm
    max}={10'}$ against their $r$-band magnitude for AKARI
  (red) and IRAS (blue).  Solid and open symbols indicate the flux
  corresponding to the single and clustering term, respectively.}
\label{fig:IRAS-AKARI}
\begin{center}
    \FigureFile(100mm,100mm){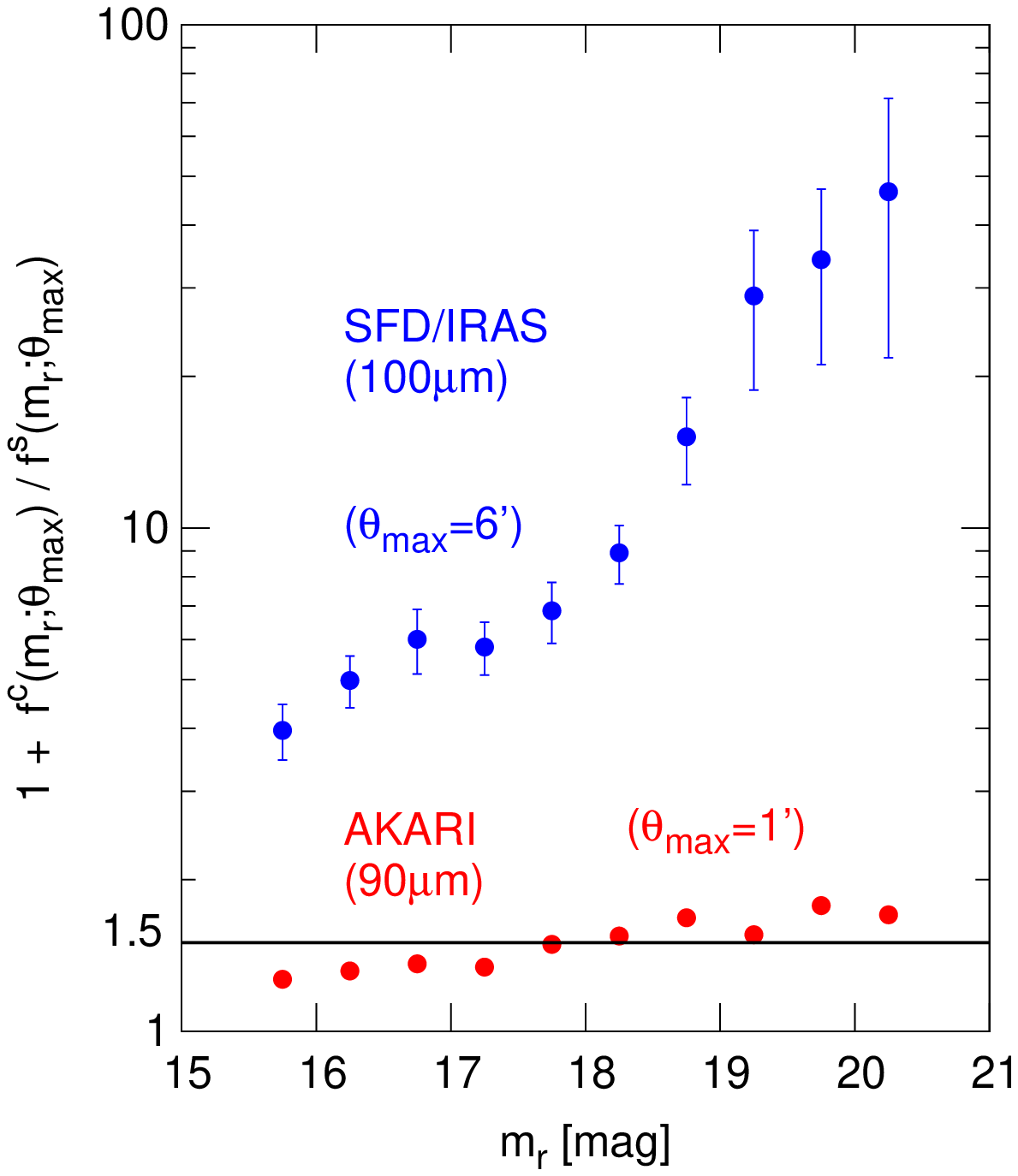}
 \end{center}
\caption{The ratio of the total flux (=$f^{\rm s}+f^{\rm c}$) to the single flux, $f^{\rm s}$, 
for both IRAS (blue) and AKARI (red).
The ranges of the integral are adopted $\theta_{\rm max}=\timeform{1.'0}, \timeform{6.'0}$ for AKARI and IRAS, respectively.}
\label{fig:total-single}
\end{figure*}
\clearpage

Figure \ref{fig:IRAS-AKARI} compares the central flux amplitudes for the
single and clustering terms ($\theta_{\rm max}=\timeform{10'}$)
estimated from the AKARI and IRAS stacking analysis.  The blue and red
symbols indicate the fluxes derived from the IRAS ($100\mum$) and AKARI
(90$\mum$), respectively.  Note that the fluxes of IRAS and AKARI are
obtained for different wavelength.  For the single term, the fluxes
derived from the IRAS and AKARI agree well except for the fainter
magnitude samples ($m_r > 18.0$).  The resulting mean $90\mum$ flux
of a galaxy of $m_r$ is well fitted to 
\begin{eqnarray}
\label{eq:f-m_r}
f^s_{90\mum}= 
13\times 10^{0.306(18-m_r)}{\rm [mJy]}.
\end{eqnarray}

On the other hand, the IRAS analysis for the clustering term
systematically over-predicts that expected from AKARI.  This should be
ascribed to the difficulty of the decomposition from the lower
angular-resolution data of IRAS.  Although the amplitudes of the
clustering fluxes depend on $\theta_{\rm max}$, the ratio of fluxes
measured by AKARI and IRAS is independent of $\theta_{\rm max}$.

Incidentally,  the presence of the clustering term tends to overestimate the flux of a galaxy if not properly decomposed. 
Figure \ref{fig:total-single} shows the overestimate factor, ($1+f^c(m_r;\theta_{\rm max})/f^s(m_r;\theta_{\rm max})$), against $m_r$ 
for both the IRAS and AKARI results. 
Here, we adopt the FWHM of PSF, $\timeform{1.'0}$ for AKARI and $\timeform{6.'0}$ for IRAS, 
as $\theta_{\rm max}$ for computing the fluxes according to equation (\ref{eq:flux-single}) and (\ref{eq:flux-clustering}). 
Therefore, this ratio basically indicates to what extent the flux of single galaxy is overestimated, 
due to the confusion of the fluxes from neighboring galaxies within the angular resolution scales of the PSFs.  
Thanks to the higher angular-resolution, the overestimation is significantly reduced for AKARI, 
but still 50\% systematic overestimate exist for $m_r>18$, 
thus we note that the fluxes of such faint galaxies are 
potentially overestimated by this level.

\begin{figure*}[h]
\begin{center}
    \FigureFile(100mm,100mm){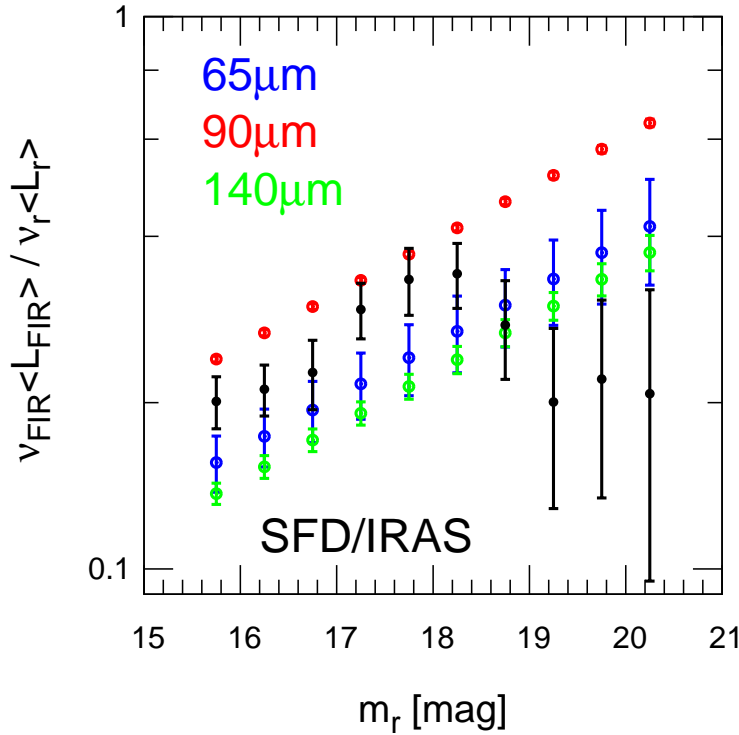}
 \end{center}
\caption{The relation between far-infrared luminosities of the central
galaxy and their $r$-band luminosities as a function of $m_r$.}
\label{fig:nuLnu}
\end{figure*}

Finally, we derive the relative luminosity at frequencies of $\nu_{\rm
FIR}$ and $\nu_{r}$ relation between $\nu_{\rm FIR}\langle L_{\rm
FIR}\rangle $ and $\nu_{r}\langle L_{r}\rangle$, since the $r$-band
magnitude, $m_r$, and the mean flux, $f^s(m_r)$, should correspond to
the same galaxy.  Figure\ref{fig:nuLnu} plots $\nu_{\rm FIR}\langle
L_{\rm FIR}\rangle / \nu_{r}\langle L_{r}\rangle $ as a function of
$m_r$.  The blue, red, and green circles indicate the ratio at 65$\mum$,
$90\mum$, and $140\mum$, respectively.  This ratio monotonically
increases as $m_r$, while KYS13 suggests that this ratio is
approximately constant from IRAS data independent of the $r$-band
magnitude (the black symbols in figure\ref{fig:nuLnu}).  The differences
of $m_r$ dependence of the luminosity ratio should be ascribed to the
underestimation of the flux for the fainter galaxies, $m_r>18.0$ (see
figure \ref{fig:IRAS-AKARI}).  Indeed, the ratio of IRAS indicates the
similar trend of AKARI for the brighter galaxies, $m_r<18.0$, where the
flux estimation is well performed.  The ratio of $\nu_{\rm FIR}\langle
L_{\rm FIR}\rangle $ to $\nu_{r}\langle L_{r}\rangle $ for $90\mum$ is
described as follows;
\clearpage
\begin{eqnarray}
\label{eq:nuL-m_r}
\frac{\nu_{\rm FIR}\langle L_{\rm FIR}\rangle }{\nu_{r}\langle L_{r}\rangle } = 
0.39\times 10^{0.096(m_r-18)}, 
\end{eqnarray}
While this increasing trend may indicate that optically faint
galaxies are relatively brighter in far-infrared, we do not discuss the
implications here since they are beyond the scope of this paper.

\section{Comparison of the clustering term in FIR and the prediction from 
the angular-correlation function of SDSS galaxies
\label{sec:Sigmac0model}}

In the last section, we estimate $\Sigma_{\rm g, obs}^{\rm c0}(m_r)$ from the
best-fit to the observed profiles of the stacking images. It is
interesting to compare the resulting value to the prediction from
equation (\ref{eq:sigmac0}), $\Sigma_{\rm g, model}^{\rm c0}(m_r)$. Indeed
KYS13 carried out the comparison and suggested that the FIR fluxes for the
clustering term obtained from the IRAS stacking analysis exceed those
predicted from the angular-correlation of the SDSS galaxies by a factor of a few. 
KYS13 speculate that this deficit may point to the existence of unknown FIR
objects other than SDSS galaxies.

Since this has very important implications, we repeat the same analysis
but with the values estimated from AKARI. As indicated in figure
\ref{fig:IRAS-AKARI}, the amplitudes of the clustering term derived from
the current AKARI analysis are systematically smaller than those from
IRAS. Thus the deficit discovered by KYS13 needs to be examined carefully
again.

For this purpose, we first note that equation (\ref{eq:sigmac0}) 
is written down more explicitly as
\begin{eqnarray}
\label{eq:sigmac0-infty}
\Sigma_{\rm g,model}^{\rm c0}(m_r;-\infty<m'<\infty) 
&=& 
2\pi\left(A\sigma_1^{2-\gamma}+(1-A)\sigma_2^{2-\gamma}\right)\left(\frac{\varphi_0}{\sqrt{2}}\right)^{\gamma}
\Gamma\left(1-\frac{\gamma}{2}\right) \cr
&{~~}&\times\int^{\infty}_{-\infty} dm' \Sigma_{\rm g}^{\rm s0}(m')
K(m',m_r) n_{\rm g}(m').
\end{eqnarray}

Since both KYS13 and our work use SDSS galaxies with $m_r=15.5$-$20.5$
in the image stacking, we can reliably compute the integral in equation
(\ref{eq:sigmac0-infty}) only for $15.5<m'<20.5$:
\begin{eqnarray}
\label{eq:sigmac0-sum}
\Sigma_{\rm g,model}^{\rm c0}(m_r;15.5<m'<20.5) 
&=&  2\pi\left(A\sigma_1^{2-\gamma}+(1-A)\sigma_2^{2-\gamma}\right)
\left(\frac{\varphi_0}{\sqrt{2}}\right)^{\gamma}
\Gamma\left(1-\frac{\gamma}{2}\right)  \cr
&{~~}&\times\int^{20.5}_{15.5} dm'\Sigma_{\rm g}^{\rm s0}(m')
K(m',m_r) n_{\rm g}(m').
\end{eqnarray}

\begin{figure*}[bt]
\begin{center}
   \FigureFile(100mm,100mm){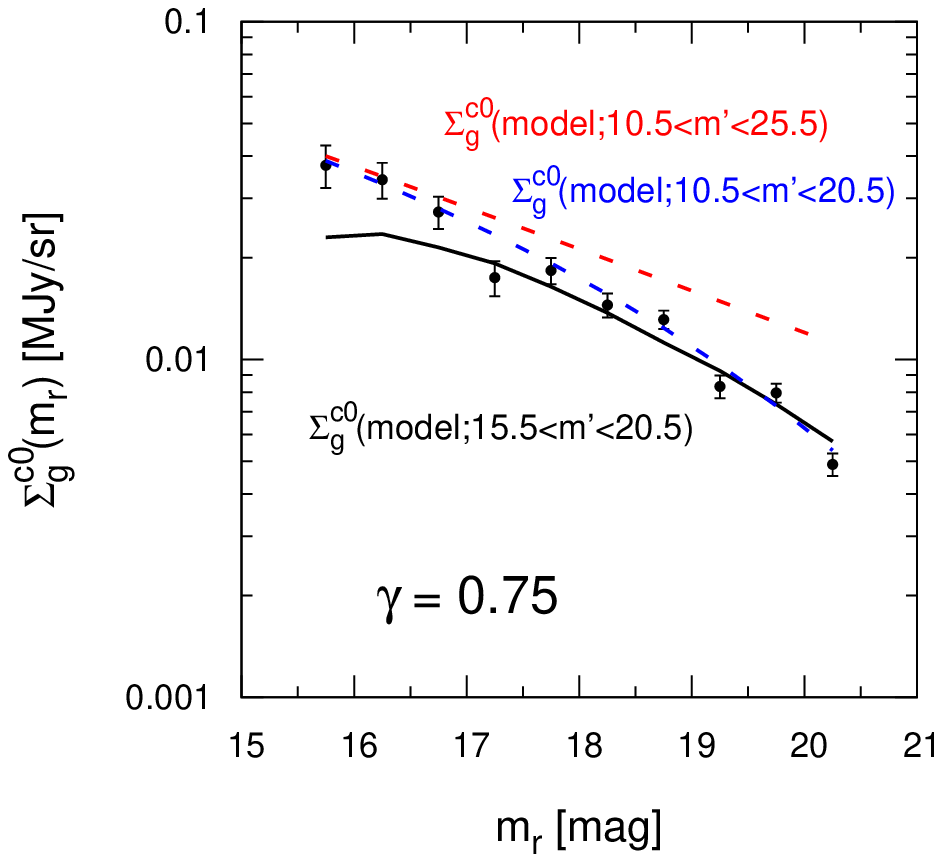}
\end{center}
\caption{The values of clustering term against their $r$-band
  magnitude. The symbols, blue dashed and red solid line indicate
  best-fit parameters from fitting, model from SDSS data $m_{r,\rm
    min}=15.5$, $m_{r,\rm min}=10.5$, respectively.}
 \label{fig:Sigmac0-rmag}
\begin{center}
    \FigureFile(100mm,100mm){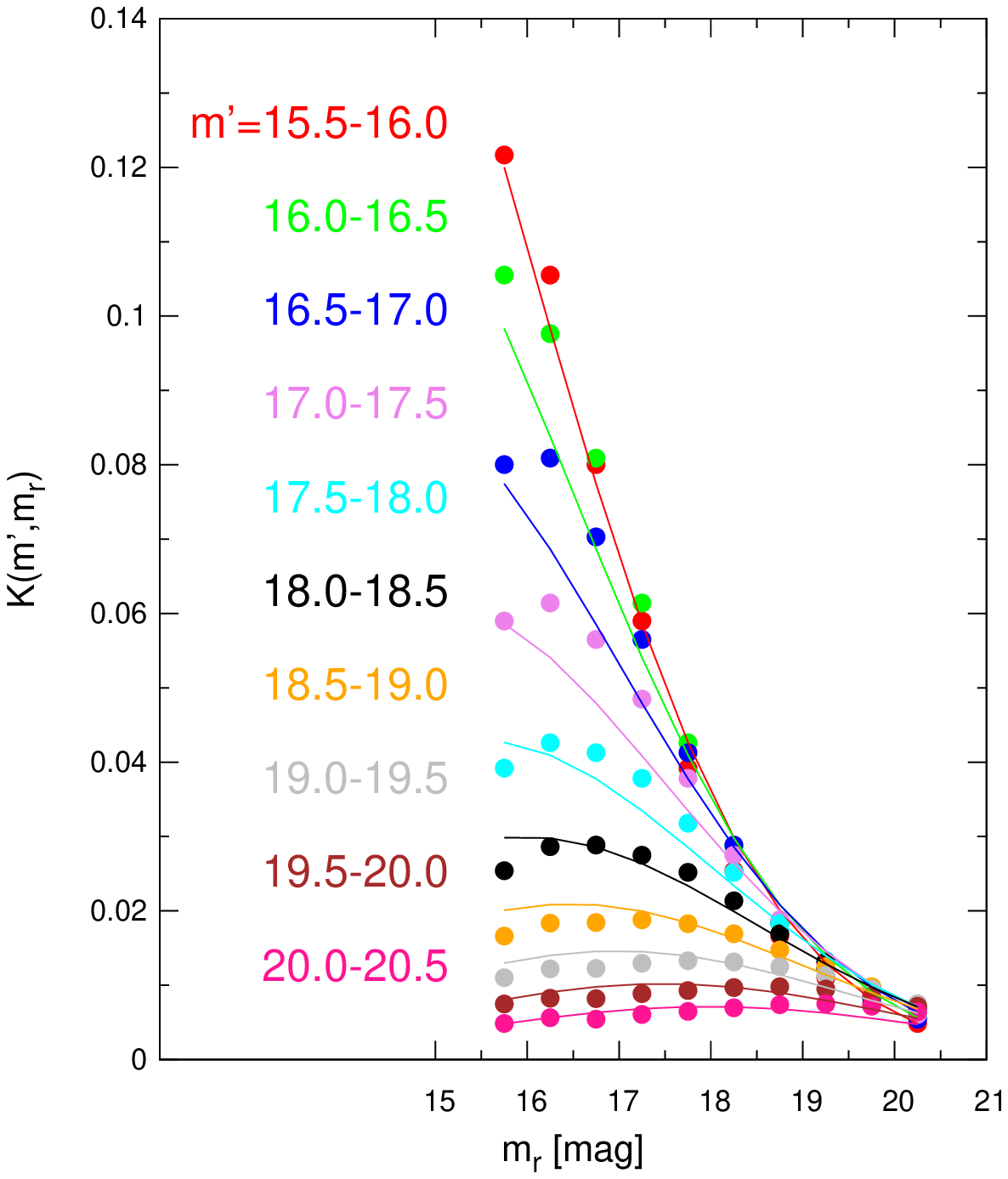}
\end{center}
\caption{Amplitudes of the angular-correlation function of SDSS
  galaxies, $K(m',m_r)$, against $m_r$.  The solid lines indicate the
  best-fit models of equation (\ref{eq:Kmodel}).}
\label{fig:K}
\end{figure*}

In figure \ref{fig:Sigmac0-rmag}, we plot the above result for
$\gamma=0.75$ as a black solid line, which should be compared with
$\Sigma_{\rm g,obs}^{\rm c0}(m_r)$ (filled circles with error-bars).
Although an overall factor of a few discrepancy in KYS13's analysis (see
their figure 10) is significantly reduced, $\Sigma_{\rm g,model}^{\rm
c0}(m_r)$ is smaller than $\Sigma_{\rm g,obs}^{\rm c0}(m_r)$ for $m_r<18$.
In this sense, our result is still consistent with that of KYS13.

Nevertheless the deficit may simply come from the upper and lower limits
of the integral in equation (\ref{eq:sigmac0-sum}).  In order to
evaluate equation (\ref{eq:sigmac0-infty}), and we extrapolate the
functions $n_{\rm g}(m_r)$, $\Sigma_{\rm g}^{\rm s0}(m_r)$ and
$K(m_r,m')$ beyond the magnitude range of the SDSS sample.

For this purpose, we use the observed $n_{\rm g}(m_r)$, $\Sigma_{\rm
 g}^{\rm s0}(m_r)$ and $K(m_r,m')$ for $15.5\leq m_r\leq20.5$, plotted
in figures \ref{fig:dndm}, \ref{fig:bestfit}, and
  \ref{fig:K}, and attempt the following power-law fits:
\begin{eqnarray}
\label{eq:dNdmmodel}
n_{\rm g}(m_r) &=& N\times10^{\nu m_r}, \\
\label{eq:sigmas0model}
\Sigma_{\rm g}^{\rm s0}(m_r) &=& S\times10^{-\mu m_r}, \\
\label{eq:Kmodel}
K(m_r,m') &=& K\times10^{-\alpha(m_r+m')-\beta(m_r-m')^2}.
\end{eqnarray}
We estimate the constants, $\alpha, \beta, \mu, \nu$, $S$, $N$, and $K$
 from the least-square fit for $\log K(m_r,m')$, $\log n_{\rm g}(m_r)$
 while $\chi^2$ minimizing fit for $\Sigma_{\rm g}^{\rm s0}(m_r)$.
The resulting fits are plotted as solid curves in figures \ref{fig:dndm},
  \ref{fig:bestfit}, and \ref{fig:K}.

We evaluate equation (\ref{eq:sigmac0-infty}) assuming that the above
extrapolation is valid.  Red and blue dashed lines in figure
\ref{fig:Sigmac0-rmag} show $\Sigma_{\rm g,model}^{\rm c0}(m_r;m_{r,\rm
min}<m'<m_{r, \rm max})$ with $m_{r,\rm min}=10.5$ and $m_{r,\rm
max}=20.5$ and $25.5$, respectively.  The result suggests that
$\Sigma_{\rm g,model}^{\rm c0}(m_r)$ sensitively depend on the
contribution from galaxies outside the observed magnitude range;
$\Sigma_{\rm g,model}^{\rm c0}(m_r=15.5)$ and $\Sigma_{\rm g,model}^{\rm
c0}(m_r=20.5)$ are significantly affected by galaxies with $m'<15.5$ and
with $m'>20.5$, respectively.  We confirmed that $\Sigma_{\rm
g,model}^{\rm c0}(m_r)$ for $15.5<m_r<20.5$ is converged if we set
$m_{r,\rm min}=10.5$ and $m_{r,\rm max}=25.5$.

Therefore the conclusion of KYS13 that the SDSS galaxies explain less than
half of the observed FIR amplitude for the clustering term is most
likely due to the magnitude limit of the SDSS galaxies themselves.
Indeed if our extrapolation model is correct, the conclusion would be
completely opposite; {\it the observed FIR fluxes are smaller than those
predicted from the SDSS galaxies}.

\clearpage

We do not take this possible discrepancy seriously, partly because it is
crucially dependent on the extrapolation of the observed
angular-correlation function, and also partly because that the galaxy
morphology and redshift evolution are totally neglected here. Since dust
emission from spirals is stronger than that from ellipticals, the
stacking analysis incorporating the morphological dependence is needed
to address the above problem properly. We attempt to qualitatively
consider the effect of the morphology dependence of $\Sigma_{\rm
g,model}^{\rm c0}(m_r)$ below.

  Just for simplicity, we assume that galaxies are divided into only two
different morphological classes; ellipticals and spirals.  We write
$n_{\rm g}(m_r)=n_{\rm e}(m_r)+n_{\rm s}(m_r)$, where indexes of the
each value e, and s correspond to ellipticals, and spirals,
respectively.  The angular-correlation functions are assumed to follow
the same power-law but with different amplitudes; $w_{\rm
ij}(m',m_r)=K_{\rm ij}(m',m_r)(\varphi/\varphi_0)^{-\gamma}$ (i,j=e,s),
where $w_{\rm ee}$, $w_{\rm es}$, and $w_{\rm ss}$ denote 
the angular-correlation functions between ellipticals and ellipticals, ellipticals
and spirals, and spirals and spirals, respectively.
In addition, we assume that
the amplitudes of the angular-correlation functions are related to each
other in terms of the constant linear bias factor as
\begin{eqnarray}
&K&_{\rm ee}(m',m_r)=b^2_{\rm e}K(m',m_r) \nonumber \\
&K&_{\rm es}(m',m_r)=K_{\rm se}(m',m_r)=b_{\rm e}b_{\rm s}K(m',m_r) \nonumber \\
\quad &K&_{\rm ss}(m',m_r)=b^2_{\rm s}K(m',m_r).
\end{eqnarray}

Then $\Sigma_{\rm g,model}^{\rm c0}(m_r)$ of equation (\ref{eq:sigmac0-infty}) is
written as
\begin{eqnarray}
\label{eq:sigmac0-model}
\Sigma_{\rm g,model}^{\rm c0}(m_r) 
&=& 
2\pi\left(A\sigma_1^{2-\gamma}+(1-A)\sigma_2^{2-\gamma}\right)
\left(\frac{\varphi_0}{\sqrt{2}}\right)^{\gamma}
\Gamma\left(1-\frac{\gamma}{2}\right) \cr
&{}& {~}\times \int dm' 
\left(\frac{n_{\rm e}(m')}{n_{\rm g}(m')}
\Sigma_{\rm g}^{\rm s0,e}(m')
+\frac{n_{\rm s}(m')}{n_{\rm g}(m')}\Sigma_{\rm g}^{\rm s0,s}(m')\right) 
\times K(m',m_r)n_{\rm g}(m') \cr
&{}& {~}\times 
\frac{n_{\rm e}(m')n_{\rm e}(m_r)b^2_{\rm e}+n_{\rm
e}(m')n_{\rm s}(m_r)b_{\rm e}b_{\rm s}+n_{\rm s}(m')n_{\rm e}(m_r)b_{\rm
e}b_{\rm s}+n_{\rm s}(m')n_{\rm s}(m_r)b^2_{s}}
{n_{\rm g}(m')n_{\rm g}(m_r)}.
\end{eqnarray}
On the other hand, the value obtained from the stacking image 
profile fitting is written as
\begin{eqnarray}
\label{eq:sigmac0-fit}
\Sigma_{\rm g,obs}^{\rm c0}(m_r) 
&=&
2\pi\left(A\sigma_1^{2-\gamma}+(1-A)\sigma_2^{2-\gamma}\right)
\left(\frac{\varphi_0}{\sqrt{2}}\right)^{\gamma}
\Gamma\left(1-\frac{\gamma}{2}\right) \cr
&{}& {}\times \int dm' 
\left[\frac{n_{\rm e}(m_r)}{n_{\rm g}(m_r)}
\left(\Sigma_{\rm g}^{\rm s0,e}(m')b^2_{\rm e}n_{\rm e}(m')
+\Sigma_{\rm g}^{\rm s0,s}(m')b_{\rm e}b_{\rm s}n_{\rm s}(m')
\right)\right. \cr
&{}& {~~~~~~~~~~~}\left.
+\frac{n_{\rm s}(m_r)}{n_{\rm g}(m_r)}
\left(\Sigma_{\rm g}^{\rm s0,s}(m')b^2_{\rm s}n_{\rm s}(m')
+\Sigma_{\rm g}^{\rm s0,e}(m')b_{\rm e}b_{\rm s}n_{\rm e}(m')
\right)\right]K(m',m_r).
\end{eqnarray}
Therefore the ratio of the integrals in equations
(\ref{eq:sigmac0-model}) and (\ref{eq:sigmac0-fit}) reduce to
\begin{eqnarray}
\label{eq:sigmac0-ratio}
r\equiv\frac{\Sigma_{\rm g,obs}^{\rm c0}}{\Sigma_{\rm g,model}^{\rm c0}} 
= \frac{(1+f_{\Sigma}f_{n}f_{b})(1+f_{n})}
{(1+f_{\Sigma}f_{n})(1+f_{n}f_{b})},
\end{eqnarray}
where
\begin{eqnarray}
\label{eq:def-stu}
   f_{\Sigma} \equiv 
\frac{\Sigma_{\rm g}^{\rm s0,e}}{\Sigma_{\rm g}^{\rm s0,s}},
\quad f_{n} \equiv \frac{n_{\rm e}}{n_{\rm s}},
\quad f_{b} \equiv \frac{b_{\rm e}}{b_{\rm s}},
\end{eqnarray}
if $f_{\Sigma}$, $f_{n}$, and $f_{b}$ are independent of $m_r$.  

For instance, if we adopt $f_{\Sigma}=0.1$ \citep{Smith2012},
$f_{n}=0.6$ \citep{Rowlands2012}, and $f_{b}=2$ \citep{Skibba2009}, we
obtain $r=0.77$. Thus our current result that $\Sigma_{\rm g,obs}^{\rm
c0}(m_r)$ is smaller than $\Sigma_{\rm g,model}^{\rm c0}(m_r)$ may be,
at least partly, accounted for by the morphological dependence of
galaxies.

For reference, we repeat the above analysis by simply splitting the
sample in two according to their $u-r$ color.  Following
\cite{Strateva2001}, we adopt $u-r = 2.22$, to separate the galaxy
types.  The number fraction of these two galaxies types is $f_{n} = 2$,
and the stacking analysis indicates $f_{\Sigma} = 0.2$.  While the
angular-correlation functions of the two types are not necessarily
described the linear bias, we take the ratio at $\theta = 10'$ and find
$f_{b} = 1.6$.  These yields the value of integral ratio, $r=0.84$.  The
morphology dependence does not seem to fully explain the possible
discrepancy plotted in figure \ref{fig:Sigmac0-rmag}.  This indicates
that the discrepancy in figure \ref{fig:Sigmac0-rmag} may be largely due
to the extrapolation of magnitude dependence of angular-correlations to
fainter magnitudes.  In any case, the reliable conclusion can be reached
only after the careful analysis in the many color bins.

\section{Mean Dust Temperature of SDSS Galaxies \label{sec:sed}}

Another important advantage of AKARI FIS is its multi-band coverage.  From the
spectrum of the stacked FIR emission in different wavelengths, we can
infer the mean dust temperature of SDSS galaxies.  \citet{Kashiwagi2015}
proposed a method to constrain the dust extent through the measurement
of dust temperature.  They obtain constraints on the dust temperature by
combining the far-infrared emission computed from the stacked images of
the IRAS map with the quasar reddening measurement by
\citet{Menard2010}.  The mean spectrum energy distribution of stacked
SDSS galaxies is a more direct probe of the dust temperature and
independent of the method proposed by \citet{Kashiwagi2015}.

We repeat the same stacking analysis over AKARI FIS maps at 65 and
140$\mum$ in addition to at 90$\mum$ as described in the previous
sections.  In particular, 140$\mum$ band is suited for the determination
of the dust temperature because it is closer to the expected peak of the
dust emission.

The best-fits of the amplitudes of the single term, $\Sigma_{\rm
g}^{\rm s0}(m_r)$, at the different bands are translated to the fluxes
$f(\nu_{\rm obs};m_r)$, and then fitted to the following grey-body spectrum:
\begin{eqnarray}
\label{eq:sed}
f(\nu_{\rm obs};m_r) = D(m_r) \nu_{\rm obs}^{3+\beta}
\left[
 \exp\left(\frac{h\nu_{\rm obs}}{k_{\rm B}T_{\rm dust,obs}}\right)
-1\right]^{-1} .
\end{eqnarray}
We adopt the dust emissivity of $\beta=1.5$ for definiteness.

\begin{figure*}[bt]
\begin{center}
    \FigureFile(100mm,100mm){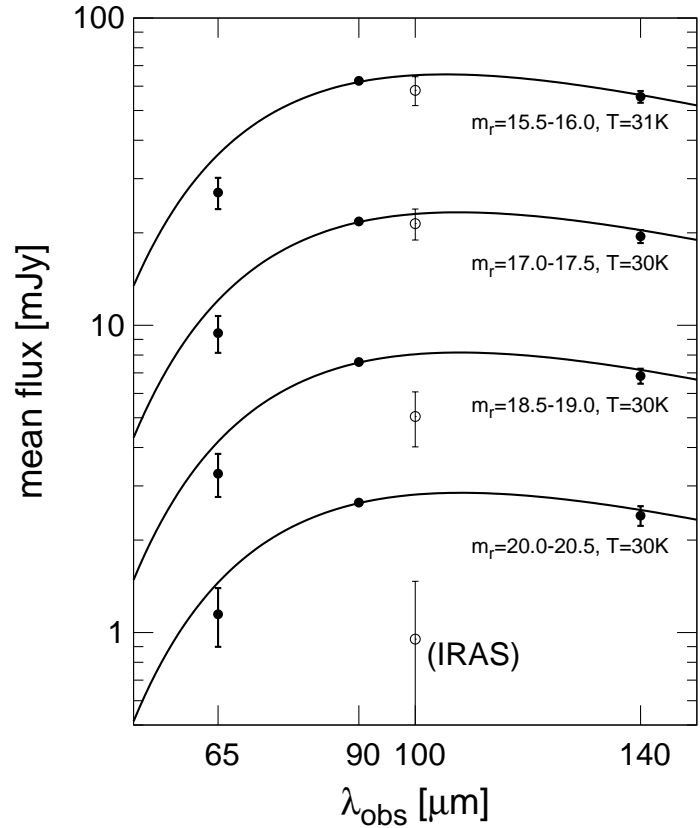}
 \end{center}
\caption{The spectral energy distribution measured by FIS.  Filled
  symbols and solid lines indicate the flux of single term measured by
  the stacking analysis (red filled symbols in
  figure(\ref{fig:IRAS-AKARI})) and their best-fits of grey-body
  (eq.(\ref{eq:sed})), respectively.  Their values of
  $m_r=15.5$-$16.0$, $17.0$-$17.5$, $18.5$-$19.0$, and $20.0$-$20.5$
  correspond to symbols and lines from top to bottom.  For reference,
  black open symbols plot the fluxes measured by the IRAS.}
\label{fig:sed}
\end{figure*}

Figure \ref{fig:sed} plots the spectral energy distribution (SED) of
galaxies with $m_r=15.5$-$16.0$, $17.0$-$17.5$, $18.5$-$19.0$, and
$20.0$-$20.5$ in filled circles.  The solid lines are their best-fits to
equation (\ref{eq:sed}), and fits are quite acceptable. The
corresponding temperatures are $T_{\rm dust,obs} \approx 31$K for
$m_r=15.5$-$16.0$, and $T_{\rm dust,obs} \approx 30$K for $17.0$-$17.5$
to $20.0$-$20.5$.  It is also encouraging that the resulting temperature
is almost independent of the magnitude as it should be.

Since the SDSS photometric galaxies are distributed over the range of
redshifts, $T_{\rm dust,obs}$ is not identical to the mean of the dust
temperature of the SDSS galaxies at different redshifts. If we assume
all the SDSS galaxies are at their median redshift $\langle z\rangle$,
the mean dust temperature at $\langle z\rangle$ is simply given by
$T_{\rm dust,obs}(1+\langle z\rangle)$. This is not an accurate but
reasonably good estimate for the redshift effect. If we adopt $\langle
z\rangle = 0.36$ \citep{Dodelson2002}, the dust temperature from our
stacking analysis is close to 40K.  

Note that we measured the dust temperature statistically, while the dust
temperature of individually identified, thus IR luminous, galaxies are
estimated with the Herschel Space Observatory (HSO)
\citep{Pilbratt2010}.  Our value is marginally consistent with 20-40K
(e.g. \citealt{Amblard2010}, \citealt{Hwang2010}, \citealt{Dunne2011},
\citealt{Hwang2011}) which are estimated with the HSO.  

Finally black open circles in figure \ref{fig:sed} are the fluxes measured
by KYS13 for IRAS. As mentioned in \S 4, the IRAS fluxes agree
well with our AKARI result, while they are systematically
underestimated for $m_r>18$.  The consistent feature is clearly
exhibited in figure \ref{fig:sed}.

\section{Summary}

 In this paper, we present the image stacking analysis of the SDSS
photometric galaxies over the AKARI Far-Infrared Surveyor maps at 65,
90, and 140$\mum$. This enables us to identify the mean FIR counterparts of
the SDSS galaxies down to the $r$-band magnitude of $m_r =20.5$ in a
statistical fashion, corresponding to 3mJy at 90$\mum$.  A typical FIR
flux of individually identified galaxies is about 10mJy.  Therefore the image
stacking analysis is useful in exploring the nature of typical galaxies
in FIR that is not easily accessible otherwise.  Our present result
improves the previous analysis using IRAS at 100 $\mum$ by
\citet{Kashiwagi2013} mainly thanks to a factor of 6 better angular
resolution of AKARI (FWHM=\timeform{53''} at 90$\mum$) relative to IRAS (FWHM=\timeform{6'} at
$100\mum$).

We decompose the stacked image profiles of galaxies with different $m_r$
into the single term and the clustering term. The single term represents
the flux from the central galaxy, while the clustering term is a sum of
contribution from near-by galaxies through their angular clustering.
Because typical sizes of SDSS galaxies are much smaller than that of
AKARI FIS PSF, the single term follows the PSF that is well approximated
by the double Gaussians. The clustering term can be written in terms of
the angular-correlation function that is directly measured from the SDSS
galaxies in optical. Thus the FIR amplitude of the clustering term
fitted from the stacked image profile can be compared with that
predicted from the SDSS data.

 This could be used in turn so as to explore the existence of the
possible unknown class of optically faint and FIR luminous objects that
are not traced by optical galaxies. Indeed \citet{Kashiwagi2013}
suggested that the FIR fluxes inferred from IRAS data are not fully
explained by the SDSS galaxies alone. Our present analysis, however,
revealed that the extrapolation of the angular-correlation function of
SDSS galaxies is sufficient to account for the detected FIR fluxes. In
other words, the discrepancy does not require any additional class of
objects, but can be explained by galaxies outside the SDSS magnitude
range.  This result may be interpreted as an example that the image
stacking analysis can put useful constraints on the angular clustering
of galaxies that are not easily probed otherwise.

We also combined the stacking analysis at 65$\mum$, 90$\mum$, and
140$\mum$, and found that the mean dust temperature of the SDSS galaxies
at $\langle z\rangle=0.36$ is $\sim 40$K. This value is slightly higher
than the typical dust temperature of galaxies that are FIR luminous and
individually detected.

Our image stacking can be improved and applied for other purposes.
First, the morphological dependence of the FIR emission of galaxies
needs to be studied as mentioned in \S \ref{sec:Sigmac0model}. Second,
the analysis based on spectroscopic, rather than photometric, samples of
galaxies is important in separating the redshift evolution and intrinsic
luminosity dependence. Third, the current method can be applied to
search for the intracluster dust that is not directly associated with
individual galaxies, in a complementary fashion to
\citet{Kashiwagi2015} and \citet{Menard2010}. Finally, it is interesting
to perform the image stacking of quasars, as has been tentatively
attempted by KYS13. We are currently working on those studies, and
plan to present the result elsewhere.

\bigskip 

\section*{Acknowledgements}

We thank Masamune Oguri and Tetsu Kitayama for useful discussion.
T.O. is supported by Advanced Leading Graduate Course for Photon Science
(ALPS) at the University of Tokyo.  T.K. gratefully acknowledges a
support from Global Center for Excellence for Physical Science Frontier
at the University of Tokyo.  This work is supported in part from the
Grant-in-Aid No. 20340041 by the Japan Society for the Promotion of
Science.  Data analysis was carried out on common use data analysis
computer system at the Astronomy Data Center, ADC, of the National
Astronomical Observatory of Japan.

This research is based on observations with AKARI, a JAXA project with
the participation of ESA.  Funding for the SDSS and SDSS-II has been
provided by the Alfred P. Sloan Foundation, the Participating
Institutions, the National Science Foundation, the U.S. Department of
Energy, the National Aeronautics and Space Administration, the Japanese
Monbukagakusho, the Max Planck Society, and the Higher Education Funding
Council for England. The SDSS Web Site is http://www.sdss.org/.

SDSS and SDSS-II are managed by the Astrophysical Research Consortium
for the Participating Institutions. The Participating Institutions are
the American Museum of Natural History, Astrophysical Institute
Potsdam, University of Basel, Cambridge University, Case Western
Reserve University, University of Chicago, Drexel University,
Fermilab, the Institute for Advanced Study, the Japan Participation
Group, Johns Hopkins University, the Joint Institute for Nuclear
Astrophysics, the Kavli Institute for Particle Astrophysics and
Cosmology, the Korean Scientist Group, the Chinese Academy of Sciences
(LAMOST), Los Alamos National Laboratory, the Max-Planck-Institute for
Astronomy (MPIA), the Max- Planck-Institute for Astrophysics (MPA),
New Mexico State University, Ohio State University, University of
Pittsburgh, University of Portsmouth, Princeton University, the United
States Naval Observatory, and the University of Washington.

\section*{Appendix. 
Validity of the subtraction method of the foreground emission  
\label{sec:sigma}}

In \S \ref{sec:results}, we subtracted the foreground emission due to
the Galactic dust, from the raw stacked images using the empirically
constructed templates for the foregrounds.  Although this method is
successful in removing the foreground contribution fairly nicely, we
carefully consider the extent to which the resulting best-fit parameters
characterizing the FIR emission of galaxies, $\Sigma_{\rm g}^{\rm
s0}(m_r)$ and $\Sigma_{\rm g}^{\rm c0}(m_r)$, are affected by this
procedure.

For that purpose, we repeated the stacking analysis of the SDSS galaxies
as presented in \S\ref{sec:stack}, but without subtracting any
foreground contribution. We model the resulting radial profiles of
galaxies as
\begin{eqnarray}
\label{eq:average-galaxy-profile-C}
\Sigma_{\rm g}^{\rm tot}({\boldsymbol \theta};m_r) =
\Sigma_{\rm g}^{\rm s}({\boldsymbol \theta};m_r) 
+\Sigma_{\rm g}^{\rm c}({\boldsymbol \theta};m_r)
+ C(m_r),
\end{eqnarray}
where $C(m_r)$ indicates the offset level due to the foreground
emission.  Here we treat $C(m_r)$ as a free parameter separately for
each magnitude bin.  We find that except that the statistical
uncertainties become slightly smaller, the best-fit values of
$\Sigma_{\rm g}^{\rm s0}(m_r)$ and $\Sigma_{\rm g}^{\rm c0}(m_r)$, are
almost identical to from those derived without subtracting the
foreground templates.  Thus, the parameters characterizing the FIR
emission of galaxies are very insensitive to how we subtract the
foreground emission.

\begin{figure}[bt]
\begin{center}
   \FigureFile(55mm,55mm){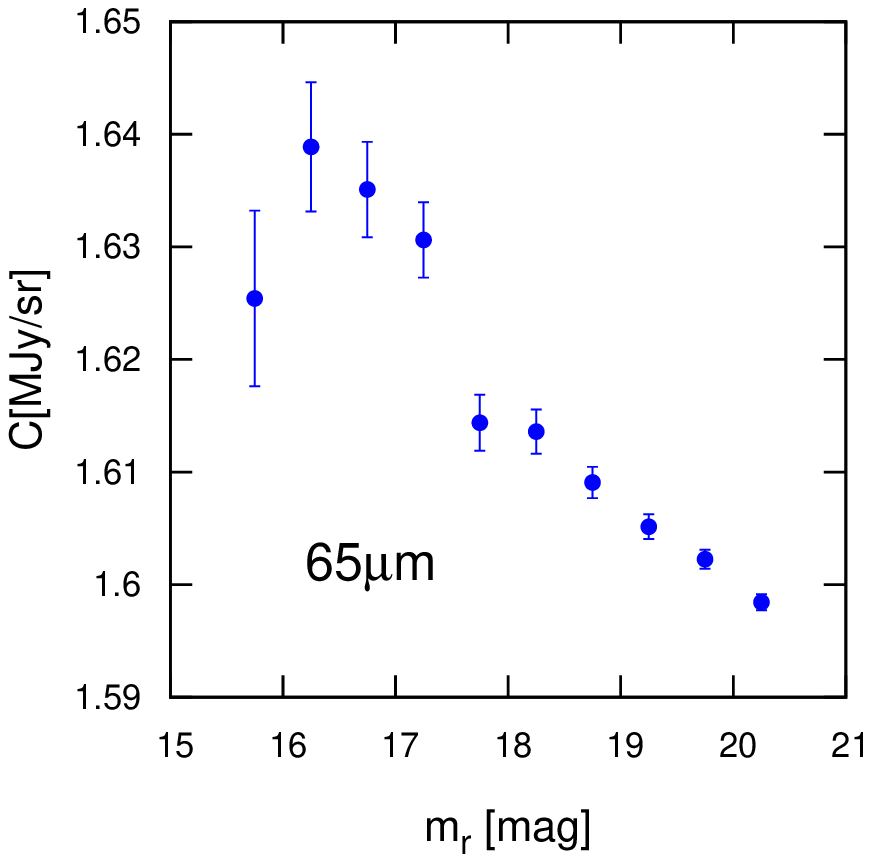}
   \FigureFile(55mm,55mm){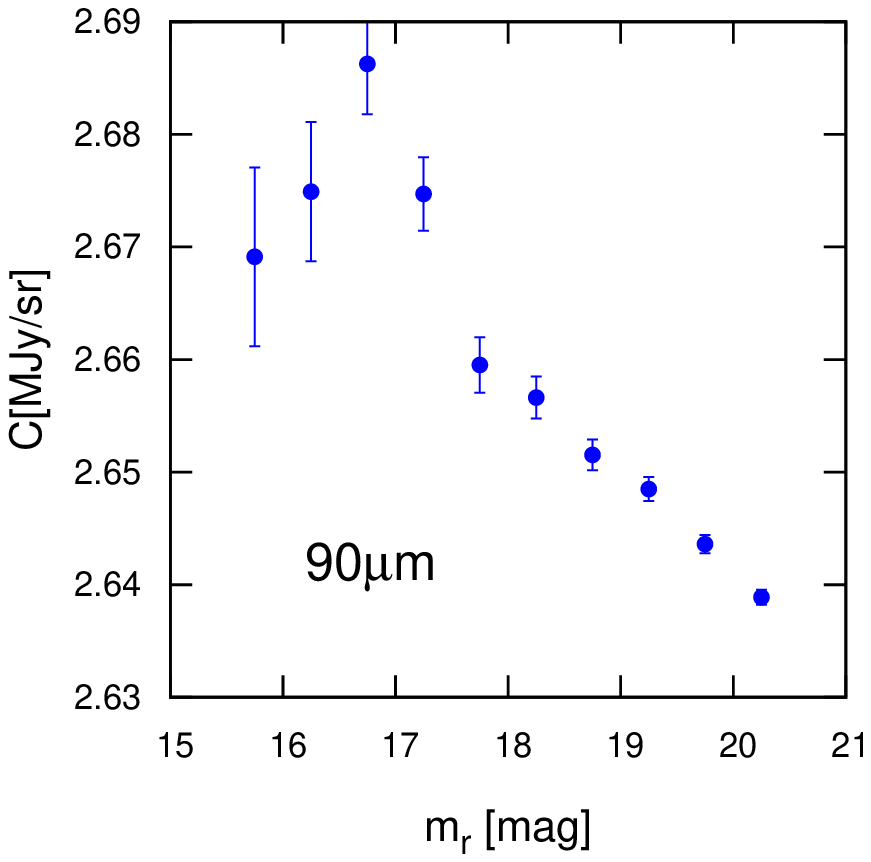}
   \FigureFile(55mm,55mm){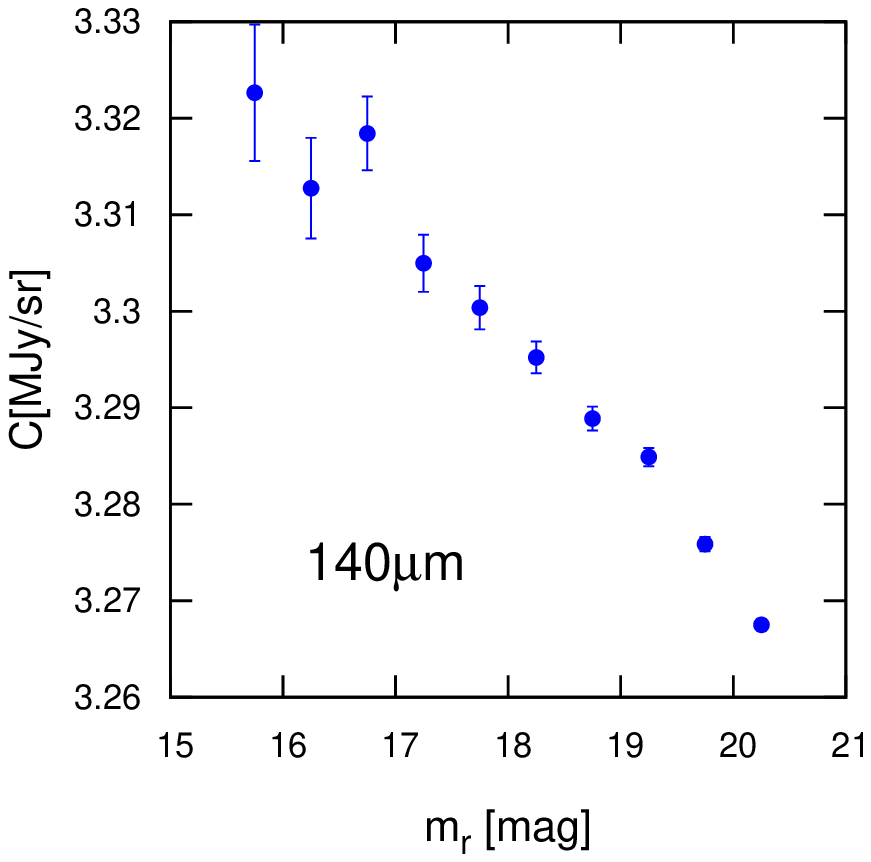}
\end{center}
\caption{The left, center, and right panels correspond to the best-fit
  parameters of constant term before removing the background at
  65$\mum$, 90$\mum$, and 140$\mum$, respectively.  } \label{fig:C}
\end{figure}

On the other hand, the best-fit parameter of $C(m_r)$ systematically
decreases against $m_r$, as shown in figure \ref{fig:C}, while the
offset level due to the foreground emission is naively expected to be
independent of $m_r$.  The similar trend was also found by KYS13, using
the IRAS data.  KYS13 suggested that the unexpected correlation is
due to the spatial inhomogeneities of the local universe, in
particular, the CfA Great Wall (hereafter, CGW).  They argued that the
CGW is coincidentally located where the IRAS 100$\mum$ emission is
relatively high.  Since the nearby structure such as the CGW would
consists of bright galaxies, the stacking results for the bright
magnitude sample could return relatively large values of the Galactic
foreground emission.

After the publication of KYS13, however, the authors found that this
explanation is not likely to be correct, due to the over-interpretation
of the results of the corresponding analysis; these results are rather
subtle and should have been interpreted more carefully, especially since
those statistical significances are limited by the number of the
galaxies used in these analysis, which is much reduced by avoiding the
CGW.

Indeed, the best-fit values of $C(m_r)$ for the AKARI results indicate
the strong correlation with $m_r$, much beyond the quoted error-bars
computed from the jackknife resampling.  Given that the jackknife
resampling method takes into account the sample variance over the survey
area, the correlation should be interpreted to be present over the
entire sky area, and would not be restricted to the specific region of
the sky area, such as the CGW.  Thus, the origin of the anomalous
correlation needs to be further investigated in detail.  We emphasize,
however, that our results for the FIR emission of galaxies would not be
affected by the value of those offset levels, as far as our assumption
that the FIR emission of galaxies basically traces the spatial
distribution of the optical galaxies.

\end{document}